\documentclass[11pt,superscriptaddress,showkeys,nofootinbib]{revtex4-1}

\usepackage{amsmath,amsthm}
\usepackage{amsfonts}
\usepackage{amssymb}
\usepackage{verbatim}
\usepackage{hyperref}
\usepackage{mathrsfs}
\usepackage{tensor}
\usepackage[dvipsnames]{xcolor}
\usepackage{newpxmath}
\usepackage[scaled=.95]{newpxtext}

\allowdisplaybreaks

\newcommand{\lin}{\\[7pt]}

\newcommand{\der}[2]{\dfrac{d#1}{d#2}}
\newcommand{\pder}[2]{\dfrac{\partial#1}{\partial#2}}
\newcommand{\pdder}[3]{\dfrac{\partial^2 #1}{\partial #2 \partial #3}}
\newcommand{\pdc}[2]{\dfrac{\partial x^{#1}}{\partial x^{#2}}}
\newcommand{\pddc}[3]{\dfrac{\partial^2 x^{#1}}{\partial x^{#2}\partial x^{#3}}}
\newcommand{\dder}[2]{\dfrac{\delta#1}{\delta#2}}
\newcommand{\pdot}[1]{\dot{\partial}_{#1}}

\newcommand{\D}{\mathcal{D}}
\newcommand{\Gd}{\mathcal{G}}
\newcommand{\Gdet}{\det{\![\mathcal{G}]}}
\newcommand{\R}{\mathcal{R}}
\newcommand{\de}{\mathrm{d}}

\newcommand{\bcg}{\,{\!}^{\text{{\tiny (b)}}}\!}

\begin{document}

\title{Weak Field Equations and Generalized FRW Cosmology on the Tangent Lorentz Bundle\footnote{This article is available for reuse under a \href{https://creativecommons.org/licenses/by-nc-nd/3.0/}{CC BY-NC-ND 3.0} license}}

 \author{A. Triantafyllopoulos}
\email{alktrian@phys.uoa.gr}
 \affiliation{Section of Astrophysics, Astronomy and Mechanics, Department of Physics, National and Kapodistrian University of Athens, Panepistimiopolis 15783, Athens, Greece}

 \author{P. C. Stavrinos}
\email{pstavrin@math.uoa.gr}
 \affiliation{Department of Mathematics, National and Kapodistrian University of Athens, Panepistimiopolis 15784, Athens, Greece}

\begin{abstract}
We study field equations for a weak anisotropic model on the tangent Lorentz bundle $TM$ of a spacetime manifold. A geometrical extension of General Relativity (GR) is considered by introducing the concept of local anisotropy, i.e. a direct dependence of geometrical quantities on observer $4-$velocity. In this approach, we consider a metric on $TM$ as the sum of an h-Riemannian metric structure and a weak anisotropic perturbation, field equations with extra terms are obtained for this model. As well, extended Raychaudhuri equations are studied in the framework of Finsler-like extensions. Canonical momentum and mass-shell equation are also generalized in relation to their GR counterparts. Quantization of the mass-shell equation leads to a generalization of the Klein-Gordon equation and dispersion relation for a scalar field. In this model the accelerated expansion of the universe can be attributed to the geometry itself. A cosmological bounce is modeled with the introduction of an anisotropic scalar field. Also, the electromagnetic field equations are directly incorporated in this framework.
\end{abstract}


\maketitle

\section{Introduction}
During the last decade there has been a considerable interest in the study of applications of Finsler geometry in different topics of Physics, such as in modified gravity theories, modern cosmology, quantum gravity e.t.c \cite{stavrinos 2004,stavrinos 2012,stavrinos-kouretsis-stathakopoulos 2008,kostelecky 2011,pfeifer-wohlfarth 2011,gibbons-gomis-pope 2007,stavrinos-diakogiannis 2004,voicu 2017,hohmann-pfeiffer 2017,kouretsis-stathakopoulos-stavrinos 2010a,stavrinos-ikeda 2006,stavrinos 2009,kostelecky 2012,vacaru 2012b,stavrinos-vacaru-vacaru 2014,vacaru 2012a,foster-lehnert 2015,fuster-pabst 2016,stavrinos-vacaru 2013,skakala-visser 2011,silva-maluf-almeida 2017,vacaru 2010a,vacaru 2011a,chang-li-wang 2013,torrome-piccione-vitorio 2012,pfeifer-wohlfarth 2012,kouretsis-stathakopoulos-stavrinos 2012a,basilakos-stavrinos 2013,papagiannopoulos-basilakos-paliathanasis-savvidou-stavrinos 2017,wang-meng 2017,lin 2013,minguzzi 2014}. It has been proposed that  Finsler gravity can be used towards studying the physical phenomena in the universe.

The development of research for the evolution of the universe can be combined with a locally anisotropic structure of the Finslerian gravitational field. Finsler-gravity models allow intrinsically local anisotropies including vector variables $y^i=\frac{dx^i}{d\tau},\, (i=0,1,2,3)$ in the framework of a tangent (Lorentz) bundle \cite{stavrinos-vacaru 2013,stavrinos-vacaru-vacaru 2014,stavrinos-ikeda 1999,pfeifer-wohlfarth 2012}. Those approaches were elaborated as unified descriptions and modifications/generalizations of Einstein gravity theory. The $y-$dependence essentially characterizes the Finslerian gravitational field and has been combined with the concept of anisotropy and the broken Lorentz symmetry which causes the deviation from Riemannian geometry, since the latter can't explain all the gravitational effects in the universe. Therefore, the consideration of Finsler geometry as a candidate for studying gravitational theories provides that matter dynamics take place \cite{silva-maluf-almeida 2017,papagiannopoulos-basilakos-paliathanasis-savvidou-stavrinos 2017}.

In the theory of Finslerian gravitational field a peculiar velocity field is produced by the gravity of mass fluctuations which are due to anisotropic distribution and motion of particles. This can be physically described in the framework of Finsler-like geometrical structure of spacetime. Einstein Finsler-like gravity theories are considered as natural candidates for investigation of local anisotropies and the dark energy problem \cite{stavrinos 2004,stavrinos-kouretsis-stathakopoulos 2008,kouretsis-stathakopoulos-stavrinos 2010a,basilakos-kouretsis-saridakis-stavrinos 2013,basilakos-stavrinos 2013,chang-li 2008,chang-li 2009,voicu 2011,vacaru 2011b}. Also, extended modified gravity theories in the framework of tangent Lorentz bundles allow generalizations of the $f(R,T,\ldots)$ ones \cite{stavrinos-vacaru 2013,stavrinos-vacaru-vacaru 2014,vacaru 2013}.

Finsler geometry gives a metric extension of the background metric of space-time in higher-order dimensions. Finsler gravity and cosmology models were developed in \cite{stavrinos-kouretsis-stathakopoulos 2008,kostelecky 2011,pfeifer-wohlfarth 2011,gibbons-gomis-pope 2007,kouretsis-stathakopoulos-stavrinos 2010a,stavrinos-diakogiannis 2004,vacaru 2012a,pfeifer-wohlfarth 2012,caponio-stancarone 2016} extending geometrical and physical ideas and were related to quantum gravity and modified dispersion relations, broken Lorentz symmetry, nonlinear symmetries and gravitational waves \cite{kouretsis-stathakopoulos-stavrinos 2010a,fuster-pabst 2016,stavrinos 2012,vacaru 2011b,kostelecky-russel 2010}. Causality on a tangent bundle is induced by Lorentzian structure of the base spacetime manifold \cite{caponio-javaloyes-sanchez 2011,ishikawa 1981,minguzzi 2014}.

Spacetimes which are described by Finsler geometry allow deviation from Lorentz invariance symmetries \cite{kouretsis-stathakopoulos-stavrinos 2010a,kouretsis-stathakopoulos-stavrinos 2010b}. A theory which naturally describes Lorentz violation phenomena in quantum gravity while preserving Einstein's general relativity in the background level is the Standard Model Extension theory (SME) \cite{kostelecky 2012,colladay-kostelecky 1997,colladay-kostelecky 1998,foster-lehnert 2015,bogoslovsky 1977,cohen-glashow 2006}. This theory is related with experimental investigations and observational efforts in astrophysics, cosmology and high energy physics \cite{bogoslovsky 2007,kostelecky-russell 1999a,kostelecky-russell 2011a,colladay-kostelecky 1997,colladay-kostelecky 1998}. In this framework Finsler structures for $b-$spaces were developed, giving a remarkable geometrization in the study of elementary particle theories \cite{foster-lehnert 2015,kostelecky 2011,kostelecky 2012,colladay-kostelecky 1997,colladay-kostelecky 1998,kostelecky-russell-tso 2012}. Additionally, the ticking rate of clocks which is crucial to the magnitude redshift calculations is determined from the background metric geometry of spacetime. In a direction-dependent space-time the ticking rate depends on the direction \cite{kostelecky 2011,kostelecky 2012,hohmann-pfeiffer 2017}.

Einstein-Finsler theories of gravity play an important role in the resolution of cosmological problems and generalize cosmological models. Based on such an approach we can get additional information for the gravity e.g. in connection with an electromagnetic field, inflaton, scalar field or spinor field. Particularly, the dynamics of Finsler geometry (velocity space) contributes decisively in the stability and acceleration of the universe. It is possible that this consideration can be used to analyze the implications of quantum gravity and related Lorentz violations in the early universe and in present day cosmology.

In the framework of Finsler extensions of general relativity, Raychaudhuri equations and energy conditions have also been studied \cite{stavrinos 2012,minguzzi 2015,stavrinos-alexiou 2016,singh-chaubey-singh 2015,mohseni-fathi 2016,thompson-fathi 2017}.

Finsler geometry includes torsions, more than one covariant derivatives and anisotropic curvatures extending the framework of field equations of general relativity and cosmology. A unified description  of the Finslerian gravitation of a spacetime manifold is given by a metric function $F$, a total metric $\Gd$ on the tangent bundle of $M$, a metrical compatible connection and a nonlinear connection $N$ \cite{vacaru-stavrinos-gaburov-gonta 2006,bucataru-miron 2007,miron-watanabe-ikeda 1987}.

This paper is organized as follows. In the second section we present the basic geometrical structures on the tangent bundle $TM$ of a manifold $M$.

In the third section a specific type of distinguished ($d-$)metric is introduced on $TM$, which consists of a background h-Riemannian perturbed by a locally anisotropic weak field. Based on a metrical compatible $d-$connection and the corresponding field equations on the tangent Lorentz bundle, we present the field equations for our model. The extra terms of the derived field equations are connected with the anisotropic sectror of geometric structure of space-time and give an interpretation for possible anisotropies of the universe.

In the fourth section we present a generalization of the definition of canonical momentum for our weak field model. Consequently, mass shell relation is also extended, and a generalization of the Klein-Gordon equation for a massive scalar field is derived. Aditionally, a modified dispersion relation for the scalar field is calculated.

In the fifth section an extended FRW cosmological model on the tangent bundle is presented in which Raychaudhuri equations, energy conditions and cosmological bounce are studied. As well, an anisotropic scalar field is introduced on $TM$ and its dynamics is described in a specific case of the weak field model. 

In the sixth section, Maxwell equations are generalized on the framework of the tangent Lorentz bundle, particularly for the weak field model.

Finally, in the concluding remarks, we summarize and discuss our results.


\section{Preliminaries}\label{sec: preliminaries}

In this section we present in brief the basic concepts of geometry on the tangent bundle of a background manifold, for more details see \cite{bucataru-miron 2007,vacaru-stavrinos-gaburov-gonta 2006,kandatu 1966,miron-watanabe-ikeda 1987,crasmareanu 2012}.

We consider a $4-$dimensional spacetime manifold $M$ and its $8-$dimensional tangent bundle $(TM,\pi,M)$ or for short $TM$, which around a point $p\in TM$ is equipped with local coordinates $\{\mathcal{X}^A \} = \{x^i,y^a\}$ where $x^i$ are the local coordinates on the base manifold $M$ around $\pi(p)$ and $y^a$ are the coordinates on the fiber. The range of values for the indices is $i,j,\ldots,s = 0,\ldots,3$ and $a,b,\ldots,f = 0,\ldots,3$.

A  local coordinate transformation is given by the relation
\begin{equation}\label{coordinate transformation x}
x^{i'} = x^{i'}(x^0,\ldots,x^3)
\end{equation}
and
\begin{equation}\label{coordinate transformation y}
y^{a'} = \pder{x^{a'}}{x^a}y^a
\end{equation}
where $x^a = \delta^a_i x^i$, $ \delta^a_i $ is the Kronecker delta, and
\begin{equation}\label{nondegenerate transformation}
\det \left\| \pder{x^{i'}}{x^j}\right\| \neq 0
\end{equation}

A nonlinear connection $N$ with coefficients $ N^a_j(x,y) $ is defined a priori on $TM$ \cite{bucataru-miron 2007,kandatu 1966,crasmareanu 2012}. Under a local coordinate transformation, coefficients $N^a_j$ obey the following transformation rule:
\begin{equation}
N^{a'}_{i'}(x,y) = \pdc{a'}{a}\pdc{i}{i'}N^a_i(x,y) + \pdc{a'}{a}\pddc{a}{b'}{i'}y^{b'} \label{nl transformation 2}
\end{equation}

On the tangent space $T_pTM$ an adapted to local coordinates basis or $ \{\delta_i,\dot\partial_a\} $ is defined by the relation
\begin{equation}
\delta_i = \dfrac{\delta}{\delta x^i}= \pder{}{x^i} - N^a_i\pder{}{y^a} \label{delta x}
\end{equation}
and
\begin{equation}
\dot \partial_a = \pder{}{y^a}
\end{equation}
For simplicity the adapted to local coordinates basis will hereafter be called adapted basis. The horizontal distribution or h-space of $T_pTM$ is spanned by $\delta_i$, while the vertical distribution or v-space of $T_pTM$ is spanned by $\pdot a$. Under a local coordinate transformation, adapted basis vectors transform as:
\begin{equation}\label{h basis transformation}
\delta_{i'} = \pder{x^i}{x^{i'}}\delta_i \qquad \pdot{a'} = \pder{x^a}{x^{a'}}\pdot{a}
\end{equation}

The adapted to local coordinates dual basis of the adjoint tangent space $T_p^*TM$ is $\{\de x^i, \delta y^a\}$ with the definition
\begin{equation}
\delta y^a = \mathrm{d}y^a + N^a_j\mathrm{d}x^j \label{delta y}
\end{equation}
For simplicity the adapted to local coordinates dual basis will hereafter be called dual adapted basis. The transformation rule for $ \{\de x^i, \delta y^a\} $ is:
\begin{equation}
\de x^{i'} = \pder{x^{i'}}{x^i}\de x^i \qquad \delta y^{a'} = \pder{x^{a'}}{x^a}\delta y^a
\end{equation}
Tensor algebra can be performed in the adapted basis in the usual way.

The bundle $TM$ is equipped with a distinguished metric ($d-$metric) $\Gd$:
\begin{equation}
\mathcal{G} = g_{ij}(x,y)\,\mathrm{d}x^i \otimes \mathrm{d}x^j + h_{ab}(x,y)\,\delta y^a \otimes \delta y^b \label{bundle metric}
\end{equation}
where the h-metric $g_{ij}$ and v-metric $h_{ab}$ are defined to be of Lorentzian signature $(-,+,+,+)$. A tangent bundle equipped with such a metric will be called a tangent Lorentz bundle. Proper time $\tau$ is defined to be measured by the norm
\begin{equation}\label{proper time}
d\tau = \sqrt{-g_{ij}(x,dx)\,dx^idx^j}
\end{equation}

The distinguished connection ($d-$connection) $ {D} $ is defined as a covariant differentiation rule that preserves h-space and v-space:
\begin{alignat}{2}
{{D}}_{\delta_k}\delta_j & = L^i_{jk}(x,y)\delta_i \qquad & {{D}}_{\dot{\partial}_c}\delta_j & = C^i_{jc}(x,y)\delta_i \label{D delta j} \\[7pt]
{{D}}_{\delta_k}\dot{\partial}_b & = L^a_{bk}(x,y)\dot{\partial}_a \qquad & {{D}}_{\dot{\partial}_c}\dot{\partial}_b & = C^a_{bc}(x,y)\dot{\partial}_a \label{D partial b}
\end{alignat}
From these the definitions for partial covariant differentiation follow as usual, e.g. for $X \in T_pTM$ we have the definitions for covariant h-derivative
\begin{equation}
\tensor{X}{^A_|_j} \equiv \overset{(h)}{{D}}_j\,X^A \equiv \delta_jX^A + L^A_{Bj}X^B \label{vector h-covariant}
\end{equation}
and covariant v-derivative
\begin{equation}
X^A|_b \equiv \overset{(v)}{{D}}_b\,X^A \equiv \dot{\partial}_bX^A + C^A_{Bb}X^B \label{vector v-covariant}
\end{equation}

The curvature of the nonlinear connection is defined by
\begin{equation}\label{Omega}
\Omega^a_{jk} = \dder{N^a_j}{x^k} - \dder{N^a_k}{x^j}
\end{equation}
The components of the torsion tensor of the $d-$connection that we need are
\begin{alignat}{3}
\mathcal{T}^i_{jk} & = L^i_{jk} - L^i_{kj} \qquad & \mathcal{T}^a_{jk} & = \Omega^a_{jk} \qquad & \mathcal{T}^a_{bc} & = C^a_{bc} - C^a_{cb} \label{torsion coefficients 1}
\end{alignat}
The h-curvature tensor of the $d-$connection in the adapted basis and the corresponding h-Ricci tensor have respectively the components
\begin{align}
R^i_{jkl} & = \delta_lL^i_{jk} - \delta_kL^i_{jl} + L^h_{jk}L^i_{hl} - L^h_{jl}L^i_{hk} + C^i_{ja}\Omega^a_{kl} \label{R coefficients 1}\lin
R_{ij} & = R^k_{ijk} = \delta_kL^k_{ij} - \delta_jL^k_{ik} + L^m_{ij}L^k_{mk} - L^m_{ik}L^k_{mj} + C^k_{ia}\Omega^a_{jk} \label{d-ricci 1}
\end{align}
while the v-curvature tensor of the $d-$connection in the adapted basis and the corresponding v-Ricci tensor have respectively the components
\begin{align}
S^a_{bcd} & = \dot{\partial}_dC^a_{bc} - \dot{\partial}_cC^a_{bd} + C^f_{bc}C^a_{fd} - C^f_{bd}C^a_{fc} \label{S coefficients 2} \lin
S_{ab} & = S^c_{abc} = \pdot{c}C^c_{ab} - \pdot{b}C^c_{ac} + C^e_{ab}C^c_{ec} - C^e_{ac}C^c_{eb} \label{d-ricci 4}
\end{align}
The generalized Ricci scalar curvature in the adapted basis is defined as
\begin{equation}
\R = g^{ij}R_{ij} + h^{ab}S_{ab} = R+S \label{bundle ricci curvature}
\end{equation}
where
\begin{align}
R=g^{ij}R_{ij} \qquad
S=h^{ab}S_{ab} \label{hv ricci scalar}
\end{align}

A $d-$connection can be uniquely defined given that the following conditions are satisfied \cite{miron-watanabe-ikeda 1987}:
\begin{itemize}
\item The $d-$connection is metric compatible
\item Coefficients $L^i_{jk}, L^a_{bk}, C^i_{jc}, C^a_{bc} $ depend solely on the quantities $g_{ij}$, $h_{ab}$ and $N^a_i$
\item Coefficients $L^i_{kj}$ and $ C^a_{bc} $ are torsion free, i.e.  $\mathcal{T}^i_{jk} = \mathcal{T}^a_{bc} = 0$
\end{itemize}
We use the symbol $\mathcal D$ instead of $D$ for a connection satisfying the above conditions, and call it a canonical and distinguished $d-$connection. Metric compatibility translates into the conditions:
\begin{equation}
\overset{(h)}{\mathcal D}_k\, g_{ij} = 0, \quad \overset{(h)}{\mathcal D}_k\, h_{ab} = 0, \quad\overset{(v)}{\mathcal D}_c\, g_{ij} = 0, \quad\overset{(v)}{\mathcal D}_c\, h_{ab} = 0
\end{equation}
The coefficients of canonical and distinguished $d-$connection can be found in \cite{miron-watanabe-ikeda 1987}:
\begin{align}
L^i_{jk} & = \frac{1}{2}g^{ih}\left(\delta_kg_{hj} + \delta_jg_{hk} - \delta_hg_{jk}\right) \label{metric d-connection 1}  \\[7pt]
L^a_{bk} & = \dot{\partial}_bN^a_k + \frac{1}{2}h^{ac}\left(\delta_kh_{bc} - h_{dc}\,\dot{\partial}_bN^d_k - h_{bd}\,\dot{\partial}_cN^d_k\right) \label{metric d-connection 2}  \\[7pt]
C^i_{jc} & = \frac{1}{2}g^{ih}\dot{\partial}_cg_{hj} \label{metric d-connection 3} \\[7pt]
C^a_{bc} & = \frac{1}{2}h^{ad}\left(\dot{\partial}_ch_{db} + \dot{\partial}_bh_{dc} - \dot{\partial}_dh_{bc}\right) \label{metric d-connection 4}
\end{align}

A geodesic curve on $TM$ is defined by the equation
\begin{equation}
\der{y^a}{\tau} + 2G^a(x,y) = 0, \qquad y^i=\frac{dx^i}{d\tau} \label{bundle geodesic}
\end{equation}
where
\begin{equation}
G^a(x,y) \equiv \frac{1}{4}\tilde{g}^{ab}\left(\pdder{\mathscr K}{y^b}{x^c}y^c - \pder{\mathscr K}{x^b}\right) \label{spray coefficients}
\end{equation}
In the above relation we defined $ \mathscr K \equiv g_{ab}(x,y)\,y^ay^b,\, y^i \equiv \delta^i_a y^a,\, g_{ab}\equiv\delta^i_a\delta^j_bg_{ij} $ and $\tilde{g}_{ab} \equiv \dfrac{1}{2}\pdder{\mathscr K}{y^b}{y^a} $.
\vspace*{3pt}
An h-vector $\xi^i = dx^i/d\lambda$ represents the horizontal part of a tangent vector on $TM$. It can be timelike, null or spacelike:
\begin{itemize}
\item timelike: $g_{ij}(x,\xi)\,\xi^i\xi^j < 0$
\item null: $g_{ij}(x,\xi)\,\xi^i\xi^j = 0$
\item spacelike: $g_{ij}(x,\xi)\,\xi^i\xi^j > 0$
\end{itemize}
The curve $x^A(\lambda)$ is timelike, null or spacelike for some value $\lambda_0$ of the parameter $\lambda$ if the tangent vector $\xi(\lambda_0)$ has the corresponding property. We see that the definition of proper time rel.(\ref{proper time}) only makes sense for a timelike segment of a curve. Massive point particles in spacetime subject only to gravity are described by timelike geodesics, while massless ones are described by null geodesics.

The symmetric part of a $(0,2)$ h-tensor $A_{ij}$ is defined as
\begin{equation}\label{symmetric}
A_{(ij)} = \frac{1}{2} \left(A_{ij} + A_{ji} \right)
\end{equation}
and the antisymmetric part of $A_{ij}$ is defined as
\begin{equation}\label{antisymmetric}
A_{[ij]} = \frac{1}{2} \left(A_{ij} - A_{ji} \right)
\end{equation}

\subsection*{The pseudo-Finsler metric}

We define a function $F(x,y):{TM}\rightarrow\mathbb{R}$ for which the following properties hold:
\begin{enumerate}
 \item $F$ is continuous on $TM$ and smooth on  $ \widetilde{TM}\equiv TM\setminus \{0\} $ i.e. the tangent bundle minus the null set $ \{(x,y)\in TM | F(x,y)=0\}$ \label{finsler field of definition}
 \item $ F $ is positively homogeneous of first degree on its second argument:
  \begin{equation}
   F(x^i,ky^a) = kF(x^i,y^a), \qquad k>0 \label{finsler homogeneity}
  \end{equation}
 \item The form \begin{equation}f_{ab}(x,y) = \dfrac{1}{2}\pdder{F^2}{y^a}{y^b} \label{finsler metric} \end{equation} defines a non-degenerate matrix: \label{finsler nondegeneracy}
  \begin{equation}
   \det\left[f_{ab}\right] \neq 0 \label{finsler nondegenerate}
  \end{equation}
\end{enumerate}
A metric function $ f_{ab}(x,y) $ given by rel.\eqref{finsler metric} is called a pseudo-Finsler metric. From properties $2$ and $3$ it becomes obvious that $ f_{ab}(x,y) $\vspace*{3pt} is positively homogeneous of zero degree on its second argument. 

Function $F(x,y)$ will generally not be smooth at the null set of the tangent bundle $TM$, as is evident from condition \ref{finsler field of definition}. Thus, relations \eqref{finsler metric} and \eqref{finsler nondegenerate} are defined only in the timelike \emph{or} only in the spacelike domain of $F(x,y)$. For this reason, instead of using $F(x,y)$ to define distances on the base manifold $M$, we use a pseudo-metric tensor $g_{ij}(x,y)$ homogeneous of degree zero on $y$ which is defined everywhere on $TM$. This tensor is the horizontal part of the metric on $TM$, rel.\eqref{bundle metric}. Metric tensor $g_{ij}(x,y)$ can always be derived from a function $F(x,y)$ when restricted on the appropriate domain, specifically for $F(x,y) = \sqrt{|g_{ij}(x,y)y^iy^j|}$ in a timelike domain of $TM$ we get
  \begin{equation}
  g_{ij}(x,y) = -\delta^a_i\delta^b_j f_{ab}(x,y)
  \end{equation}
  while for the same $F(x,y)$ in a spacelike domain of $TM$ we get
  \begin{equation}
  g_{ij}(x,y) = \delta^a_i\delta^b_j f_{ab}(x,y)
  \end{equation}

From rel.\eqref{proper time} we see that the proper time
\begin{equation}
\tau = \int_{\lambda_1}^{\lambda_2}\left({-g_{ij}(x,x')\frac{dx^i}{d\lambda}\frac{dx^j}{d\lambda}}\right)^{1/2}d\lambda
\end{equation}
where $x'=dx/d\lambda$, is independent from the choice of parametrization of the path due to homogeneity of zero degree of the h-metric on $x'$.

Geodesics equation rel.(\ref{bundle geodesic}) in the case of a pseudo-Finsler metric takes the form
\begin{equation}
\der{y^a}{\tau} + \gamma^a_{ij}\,y^i y^j = 0,\qquad y^i=\frac{dx^i}{d\tau} \label{finsler geodesics}
\end{equation}
where the Christoffel symbols of the second kind for the h-metric are
\begin{equation}
\gamma^k_{ij} = \frac{1}{2}g^{kl}\left(\partial_jg_{il} + \partial_ig_{jl} - \partial_lg_{ij}\right)
\end{equation}
and
$\gamma^a_{ij} = \delta^a_k \gamma^k_{ij}$.

\section{Field equations for a weakly anisotropic model}\label{sec: weak field equations}

In this section, we study weak field equations in the framework of a tangent Lorentz bundle. Previous approaches concerning a weak field limit on the tangent bundle have been studied in \cite{balan-stavrinos 2001,balan-stavrinos 2004}.

We consider a tangent bundle $TM$ equipped with a $d-$metric
\begin{equation}
\mathcal{G}(x,y) = g_{ij}(x,y)\,\mathrm{d}x^i \otimes \mathrm{d}x^j +  h_{ab}(x,y)\,\delta y^a \otimes \delta y^b \label{total metric}
\end{equation}
where the h-metric and v-metric can be decomposed as:
\begin{equation}\label{h metric}
g_{ij}(x,y) = \bcg g_{ij}(x) + \tilde g_{ij}(x,y)
\end{equation}
and
\begin{equation}\label{v metric}
h_{ab}(x,y) = \eta_{ab} + \tilde h_{ab}(x,y)
\end{equation}
where $ \bcg g_{ij}(x) $ is the background h-space metric, $\eta_{ab}$ is the background v-space Minkowski metric, $\tilde g_{ij}(x,y)$, $\tilde h_{ab}(x,y)$ are weak tensorial anisotropic ($y$-dependent) fields with $|\!\det[\tilde g]| \ll 1$ and $ |\!\det [\tilde h]| \ll 1$, $\det [\tilde g]$, $\det [\tilde h]$ are the determinants of $\tilde g_{ij}$ and $\tilde h_{ab}$ respectively. The signature convention $(-,+,+,+)$ is assumed for the individual h-space and v-space background metrics.
With these definitions, rel.\eqref{total metric} is defined to be the sum of a background metric $\bcg \, \Gd (x)$ and a perturbation $\tilde \Gd(x,y)$, where
\begin{equation}\label{background metric}
\bcg\, \Gd(x) = \bcg g_{ij}(x)\,\de x^i \otimes \de x^j + \eta_{ab}\,\delta y^a \otimes \delta y^b 
\end{equation}
and
\begin{equation}\label{weak field}
\tilde \Gd(x,y) = \tilde g_{ij}(x,y)\,\de x^i \otimes \de x^j + \tilde h_{ab}(x,y)\,\delta y^a \otimes \delta y^b 
\end{equation} 

The specific choice of metric on $TM$ allows us to readily generalize Einstein's general relativity, resulting from rel.\eqref{h metric} being the sum of a pseudo-Riemannian space metric (in the sense that it only depends on the position $x$ on the base manifold $M$) and a weak anisotropic field. On the other hand, the v-metric of the v-space has no corresponding form in general theory of relativity. In that case, we consider the sum of a flat background and a weak anisotropic field rel.(\ref{v metric}).

The background Christoffel symbols are
\begin{equation}\label{christoffel background}
\bcg\,\gamma^i_{jk}(x) = \bcg g^{ir}(x)\,\bcg\,\gamma_{rjk}(x) = \frac{1}{2}\bcg g^{ir}(x)\, \big[ \partial_k\bcg g_{jr}(x) + \partial_j\bcg g_{kr}(x) - \partial_r\bcg g_{jk}(x)\big]
\end{equation}
and the corresponding Riemann curvature tensor, Ricci tensor and Ricci scalar curvature are
\begin{gather}
\bcg R^r_{ikj} = \partial_j\bcg\,\gamma^r_{ik} - \partial_k\bcg\,\gamma^r_{ij} + \bcg\,\gamma^m_{ik}\,\bcg\,\gamma^r_{mj} - \bcg\,\gamma^m_{ij}\,\bcg\,\gamma^r_{mk} \label{riemann background} \lin
\bcg R_{ij} = \bcg R^k_{ijk} = \partial_k\bcg\,\gamma^k_{ij} - \partial_j\bcg\,\gamma^k_{ik} + \bcg\,\gamma^m_{ij}\,\bcg\,\gamma^k_{mk} - \bcg\,\gamma^m_{ik}\,\bcg\,\gamma^k_{mj} \label{ricci background} \lin
\bcg R = \bcg g^{ij}(x) \bcg R_{ij}
\end{gather}
From relations \eqref{h metric} and \eqref{v metric} the inverse h-metric and v-metric to first order with respect to $\tilde g_{ij}$ and $\tilde h_{ab}$ immediately follow:
\begin{equation}
g^{ij}(x,y) = \bcg g^{ij}(x) - \tilde g^{ij}(x,y)
\end{equation}
and
\begin{equation}
h^{ab}(x,y) = \eta^{ab} - \tilde h^{ab}(x,y)
\end{equation}
where
\begin{equation}
\tilde g^{ij} = \bcg g^{ik}\,\bcg g^{jl}\,\tilde g_{kl}
\end{equation}
and
\begin{equation}
\tilde h^{ab} = \eta^{ac}\,\eta^{bd}\,\tilde h_{cd}
\end{equation}

\subsection{Proper time in the weak field model}

By using Taylor expansion in rel.\eqref{proper time}, proper time can be written in the weak-field as 
\begin{equation}\label{proper time weak field}
d\tau = \sqrt{-\bcg g_{ij}(x)dx^i dx^j} - \frac{1}{2}\big(-\bcg g_{ij}(x)dx^i dx^j \big)^{-1/2}\tilde g_{ij}(x,y)dx^idx^j
\end{equation}
where we kept terms up to first order with respect to $\tilde g_{ij}(x,y)$. The extra term on the rhs of \eqref{proper time weak field} can cause a change in the ticking rate of a clock depending on its orientation. Relation \eqref{proper time weak field} is generally not an invariant quantity under the Lorentz symmetry group, as a Lorentz boost will not just transform the local frame of reference, but also change the position on the fiber, on which $\tilde g_{ij}(x,y)$ depends. This is an example of Lorentz violation due to anisotropy.

\subsection{Connection coefficients and curvature of the model}

We use in the following a canonical and distinguished $d-$connection $\D$, rel.(\ref{metric d-connection 1}-\ref{metric d-connection 4}). We additionally consider the connection to be Cartan-type \cite{miron-watanabe-ikeda 1987,vacaru-stavrinos-gaburov-gonta 2006}, so we have
\begin{align}
& N^a_j = L^a_{bj}y^b \label{cartan type 1}\lin
& C^a_{bc}y^b = 0 \label{cartan type 2}
\end{align}
In the following we calculate the connection coefficients for the weakly anisotropic metric defined in relations \eqref{total metric}, \eqref{h metric} and \eqref{v metric}.

We define a weak nonlinear connection $ \tilde N^a_i $ on $TM$ homogeneous of degree $1$ on $y$ in accordance with \cite{kandatu 1966}, and demand that it is of the same order with $ \tilde g_{ij} $ and $ \tilde h_{ab} $. By using Euler's theorem on homogeneous functions we get:
\begin{equation}\label{N euler}
y^b\pdot{b} \tilde N^a_i = \tilde N^a_i
\end{equation}
From relations \eqref{cartan type 1} and \eqref{metric d-connection 2} we get to first order with respect to $\tilde h_{ab}$:
\begin{equation}
\tilde N^a_i = y^b\partial_i\tilde h^a_b - y_b\eta^{ac}\pdot{c}\tilde N^b_i \nonumber
\end{equation}
Contracting with $y_a$ gives:
\begin{equation}
y_b\left( \tilde N^b_i - \frac{1}{2}y_a \partial_i\tilde h^{ab}\right) = 0 \nonumber
\end{equation}
and since this holds for arbitrary $y$ we deduce:
\begin{equation}\label{N weak}
\tilde N^a_i = \frac{1}{2}y_b \partial_i\tilde h^{ab}
\end{equation}
Since $\tilde N^a_i(x,y)$ is homogeneous of degree $1$ on $y$, from rel.\eqref{N weak} it is deduced that $\tilde h_{ab}(x,y)$ is homogeneous of degree zero and by extension $h_{ab}(x,y) = \eta_{ab} + \tilde h_{ab}(x,y)$ is also homogeneous of degree zero.

Now we calculate to first order with respect to $\tilde h_{ab}$ the curvature coefficients of the nonlinear connection from relations \eqref{Omega} and \eqref{N weak} and we find:
\begin{equation}
\Omega^a_{jk} = 0
\end{equation}
The connection coefficients defined in rel.(\ref{metric d-connection 1}-\ref{metric d-connection 4}) give to first order with respect to $\tilde g_{ij}$ and $\tilde h_{ab}$:
\begin{align}
L^i_{jk} = & \, \gamma^i_{jk} = \bcg\,\gamma^i_{jk} - \tilde g^{ir}\,\bcg\,\gamma_{rjk} + \frac{1}{2}\bcg g^{ir}\big(\partial_k\tilde g_{jr} + \partial_j\tilde g_{kr} - \partial_r\tilde g_{jk} \big) \label{d-1 first order}\lin
L^a_{bi} = & \,\frac{1}{2}\partial_i\tilde h^a_b \label{d-2 first order} \lin
C^i_{jc} = & \,\frac{1}{2}\,\bcg g^{ik}\pdot{c}\tilde g_{kj} \label{d-3 first order} \lin
C^a_{bc} = & \,\eta^{ad}\left( \pdot{c}\tilde h_{bd} + \pdot{b}\tilde h_{cd} - \pdot{d} \tilde h_{bc}\right)  \label{d-4 first order}
\end{align}
The h-Ricci curvature coefficients and the h-Ricci scalar are given in Appendix \ref{first order Ricci appendix}.
From relations \eqref{d-ricci 4} and \eqref{d-4 first order} we get to first order with respect to $\tilde g_{ij}$ and $\tilde h_{ab}$ the v-components of the Ricci tensor:
\begin{equation}
S_{ab} = \frac{1}{2}\left( \dot{\partial}^c\pdot{a} \tilde h_{bc} + \dot{\partial}^c\pdot{b} \tilde h_{ac} - \overset{(v)}{\square} \tilde h_{ab} - \pdot{a}\pdot{b} \tilde h\right) \label{ricci s first order}
\end{equation}
where $ \tilde h \equiv \tilde h^c_c $ is the trace of the perturbation on the v-metric and $ \overset{(v)}{\square} \equiv \dot{\partial}^c\pdot{c} $ is the d'Alembertian of the background v-space. From relations \eqref{hv ricci scalar} and \eqref{ricci s first order} we get the v-space scalar curvature to first order with respect to $\tilde h_{ab}$:
\begin{equation}\label{ricci scalar s first order}
S = \pdot{a}\pdot{b} \tilde h^{ab} - \overset{(v)}{\square} \tilde h
\end{equation}
From this relation we immediately get the condition for the vanishing of the v-space curvature $S$ to first order with respect to $\tilde h_{ab}$ as
\begin{equation}
\pdot{a}\pdot{b} \tilde h^{ab} = \overset{(v)}{\square} \tilde h
\end{equation}

\subsection{Field equations}
Field equations on the tangent bundle for a distinguished connection with coefficients given in relations (\ref{metric d-connection 1}-\ref{metric d-connection 4}) are derived in Appendix \ref{sec: field equations} using the variational principle, resulting in relations \eqref{hfe} and \eqref{vfe}, which in our case are written as:
\begin{gather}
R_{(ij)} - \frac{1}{2}(R+S)g_{ij} = \kappa \overset{1}T_{ij} \label{g field equation} \lin
S_{ab} - \frac{1}{2}(R+S)h_{ab} = \kappa \overset{2}T_{ab} \label{h field equation}
\end{gather}
where
\begin{align}
\overset{1}T_{ij} \equiv -\frac{2}{\sqrt{\det [\Gd]}}\frac{\delta\left(\sqrt{\det [\Gd]}\,\mathcal{L}_M\right)}{\delta g^{ij}} \label{h-EM tensor} \lin
\overset{2}T_{ab} \equiv -\frac{2}{\sqrt{\det [\Gd]}}\frac{\delta\left(\sqrt{\det [\Gd]}\,\mathcal{L}_M\right)}{\delta h^{ab}} \label{v-EM tensor}
\end{align}
are the coefficients of the generalized energy-momentum tensor on $TM$, $ \det [\Gd]$ is the metric tensor's determinant and $ \mathcal{L}_M $ denotes the Lagrangian density of the matter fields.

We will make some comments in order to give some physical interpretation to the energy momentum v-tensor, rel.\eqref{v-EM tensor}, which is an object with no equivalent in Riemannian gravity. Lorentz violations produce anisotropies in the space and the matter sector [7][64][65]. These act as a source of anisotropy and can contribute to the energy-momentum tensors of the horizontal and vertical space $\overset{1}{T}_{ij}$ and $\overset{2}{T}_{ab}$. Energy-momentum tensor $\overset{2}{T}_{ab}$ contains more information of anisotropy which is produced from the metric  $h_{ab}$ including additional internal degrees of freedom.

From relations \eqref{g field equation}, \eqref{ricci r first order}, \eqref{ricci scalar r first order} and \eqref{ricci scalar s first order} we get the h-space field equations as:
\begin{equation}
\bcg R_{ij} - \frac{1}{2}\,\bcg g_{ij}\,\bcg R + \mathcal{A}_{ij} = \kappa \overset{1}T_{ij} \label{field equation 1 weak}
\end{equation}
From relations \eqref{h field equation}, \eqref{ricci s first order}, \eqref{ricci scalar r first order} and \eqref{ricci scalar s first order} we get the v-space field equations as:
\begin{equation}
-\frac{1}{2}\eta_{ab}\bcg R + \mathcal{B}_{ab} = \kappa \overset{2}T_{ab} \label{field equation 2 weak}
\end{equation}
where $ \mathcal{A}_{ij} $ and $ \mathcal{B}_{ab} $ are the terms of the h-space and v-space field equations which are linear in $\tilde g_{ij}$ and $\tilde h_{ab}$, see Appendix \ref{first order coefficients appendix}.

\section{Mass shell and dispersion relation}\label{sec: mass shell and dispersion}
In this section we study the dynamics of a massive point particle and we compare it with the procedure of \cite{kostelecky-russel 2010}.
\subsection{Generalized mass shell equation}\label{subsec: mass shell}
We consider a Lagrangian $L$ homogeneous of degree one on $y^i$  and the action
\begin{equation}\label{particle action}
S=\int L \, d\tau
\end{equation}
The Lagrangian $ L $ is defined as
\begin{equation}\label{pp lagrangian}
L = -m \left( -g_{ij}(x,y) \, y^i y^j\right)^{1/2},\qquad y^i=\frac{dx^i}{d\tau}
\end{equation}
where $m$ is the rest mass of the point particle and $\tau$ is the proper-time, rel.\eqref{proper time weak field}. We limit ourselves to an h-metric that can be decomposed according to rel.\eqref{h metric} and substituting to rel.\eqref{pp lagrangian} we get to first order with respect to $\tilde g_{ij}$\footnote{Post-publish remark: no assumptions on the homogeneity of $\tilde g_{ij}$ were made to derive the results of this section.}:
\begin{equation}\label{pp lagrangian first order}
L = -m \left( L_R - \frac{1}{2}L_R^{-1}\tilde g_{ij}(x,y) \, y^i y^j\right)
\end{equation}
where the Riemannian norm $L_R$ is defined as
\begin{equation}\label{Riemannian norm}
L_R = \left( - \bcg g_{ij}(x) \, y^i y^j\right)^{1/2}
\end{equation}
The canonical four-momentum is
\begin{equation}\label{4 momentum}
p_i = \pder{L}{y^i}
\end{equation}
By using the Lagrangian rel.\eqref{pp lagrangian first order} and setting $c=1$ we get
\begin{equation}\label{4 momentum first order}
p_i = m \left[ y_i + \frac{1}{2} \left(\pdot{i} \tilde g_{kl}\right) y^k y^l \right]
\end{equation}
see Appendix \ref{app: massive particle momentum} for more details.

This relation can be used to calculate the generalized mass shell equation for our framework:
\begin{equation}\label{generalized mass shell}
p^i p_i = m^2 \left[ -1 + y^i \left( \pdot{i} \tilde g_{kl}\right) y^k y^l  \right]
\end{equation}

Rel.\eqref{4 momentum first order} gives to zeroth order $p_i = m  y_i$, so rel.\eqref{generalized mass shell} can be equivalently written to first order with respect  $\tilde g_{ij}$ as:
\begin{equation}\label{generalized mass shell final}
p^i p_i = -m^2  + \frac{p^i}{m}\left(\pdot{i} \tilde g_{kl} \right)p^k p^l
\end{equation}

\subsection{Dispersion relation}\label{subsec: dispersion}
We know from Riemannian geometry that we can choose a coordinate system so that locally we get $\bcg g_{ij} = \eta_{ij} = diag(-1,1,1,1)$ and the coefficients $\bcg\,\gamma^k_{ij}$ vanish. We call this a local inertial frame. We work on such a frame, and follow the usual method for quantization of physical quantities by replacing them with operators, so in {position space} we get for the position operator $x^i \rightarrow \mathbf{x}^i = x^i$ and for the momentum operator
\begin{equation}\label{quantization}
p_j \rightarrow \mathbf{p}_j = -i\hbar\,\partial_j \Rightarrow p^i p_i \rightarrow \mathbf{p}^i \mathbf{p}_i = - \hbar^2\square = - \hbar^2 \eta^{ij} \partial_i \partial_j
\end{equation}
where we denoted quantum operators with boldface.

Perturbation $\tilde g_{ij}(x,p)$ depends on both position and momentum, so upon quantization there is an ambiguity regarding the ordering of the operators, as $\mathbf{x}^i$ and $\mathbf{p}_i$ do not commute. We will not attempt to treat this ambiguity in the present work, instead we will restrict ourselves to the case where the metric perturbation $\tilde g_{ij}$ is only $y-$dependent. This way only the momentum operator appears upon quantization, so we don't need to worry about ordering.

This approach gives a generalized equation for a scalar (spin-0) boson field $\phi(x)$:
\begin{equation}\label{klein gordon generalized}
\square \phi = \frac{m^2}{\hbar^2}\phi  - \frac{i \hbar}{m}\eta^{ij}\left( \pdot{j} \tilde{\mathbf{g}}^{kl}\right) \partial_i \partial_k \partial_l \phi
\end{equation}
where $ \tilde{\mathbf{g}}^{kl}(\partial_j) $ and $ \pdot{r} \tilde{\mathbf{g}}^{kl}(\partial_j) $ are operators that occur from $ \tilde g^{kl}(p_j) $ and $ \pdot{r}\tilde g^{kl}(p_j) $ respectively by following the procedure in rel.\eqref{quantization}. This is a generalization of the Klein-Gordon equation from the standard model of particle physics, and for $\tilde g_{ij} = 0$ we get $ \square \phi = \frac{m^2}{\hbar^2}\phi $, which is the well-known Klein-Gordon equation. We see that the extra terms in \eqref{klein gordon generalized} are due to spacetime anisotropy. 

To calculate the dispersion relation for the particle, we set
\begin{equation}\label{wavevector}
p_i = \hbar k_i = \hbar \left(- \omega_{\vec{k}}, \vec{k} \right)
\end{equation}
in accordance with the procedure of \cite{kostelecky-russel 2010}. From rel.\eqref{wavevector} and applying Eq.\eqref{generalized mass shell final} we get the dispersion relation:
\begin{align}\label{dispersion relation}
\omega_{\vec k}^2 = & \, \|\vec{k}\|^2 + \frac{m^2}{ \hbar^2} \nonumber \lin
& - \frac{\hbar}{m}\left[ \omega_{\vec{k}}^2 \left( \omega_{\vec{k}}\, \dot{\partial}^0 + k_\alpha \dot{\partial}^\alpha \right) \tilde g^{00} -2\omega_{\vec k}\, k_\beta\left( \omega_{\vec{k}}\, \dot{\partial}^0 + k_\alpha \dot{\partial}^\alpha \right) \tilde g^{0\beta} + k_\beta k_\gamma \left( \omega_{\vec{k}}\, \dot{\partial}^0 + k_\alpha \dot{\partial}^\alpha \right) \tilde g^{\beta\gamma} \right]
\end{align}
where a greek letter indicates a spatial index ($ \alpha, \beta, \ldots = 1,2,3$). This is a generalization of flat isotropic spacetime dispersion relation $ \omega_{\vec k}^2 = \|\vec{k}\|^2 + m^2/\hbar^2 $. We see that the extra terms on the rhs of \eqref{dispersion relation} are due to spacetime anisotropy.

In general relativity, the quantity $\pder{p_0}{p_\beta} $ is interpreted as the group velocity $v_{gr}$ of the waveform and is equal to $ -\dfrac{y^\beta}{y^0} $. We will study whether this equality holds in our weak field framework. We differentiate rel.\eqref{generalized mass shell final} with respect to $p_\beta$ and after straightforward calculations we get
\begin{equation}\label{momentum derivative 1}
\pder{p_0}{p_\beta} = -\frac{p^\beta}{p^0} - \frac{p^\beta}{2(p^0)^2} \pder{\tilde g_{ij}}{p_0} p^i p^j + \frac{1}{2 p^0} \pder{\tilde g_{ij}}{p_\beta} p^i p^j
\end{equation}
By using relation \eqref{4 momentum first order} we get the following relations keeping terms up to first order:
\begin{equation}\label{p to y derivative}
\pder{\tilde g_{kl}}{p_i} = \eta^{ij}m^{-1}\pdot{j} \tilde g_{kl}
\end{equation}
\begin{equation}\label{p to y ratio}
\frac{p^\beta}{p^0} = \frac{y^\beta}{y^0} + \frac{1}{2 y^0}y^k y^l \dot{\partial}^\beta \tilde g_{kl} - \frac{y^\beta}{2(y^0)^2}y^k y^l \dot{\partial}^0 \tilde g_{kl}
\end{equation}
Putting together relations \eqref{momentum derivative 1}, \eqref{p to y derivative} and \eqref{p to y ratio} we arrive at the equation:
\begin{equation}\label{group velocity}
\pder{p_0}{p_\beta} = -\frac{y^\beta}{y^0}
\end{equation}
In conclusion, using homogeneity condition for the Lagrangian \eqref{pp lagrangian first order} of our generalized space, we arrive at an extended dispersion relation \eqref{dispersion relation} which satisfies the group velocity equation \eqref{group velocity}.  Note that we did not have to assume any homogeneity on $\tilde g_{kl}$ itself to obtain this result. We observe that the method we developed is consistent with the one presented in \cite{kostelecky-russel 2010} .

\section{Generalized FRW cosmology of the model}\label{sec: generalized frw cosmology}

Cosmological evolution of the universe is described by the well-known spatially flat FRW metric. The dynamics of this metric is determined by the Friedmann equations. In order that these equations agree with the accelerated expansion of the universe suggested by various observational data, one has to assume the existence of some exotic matter field usually called dark energy. Some studies in order to explain the accelerated expansion, using only geometrical structures, involve modified theories of gravity, e.g. Finsler-Randers cosmology \cite{papagiannopoulos-basilakos-paliathanasis-savvidou-stavrinos 2017,basilakos-kouretsis-saridakis-stavrinos 2013}. In this section, we make an effort to model this acceleration using the extra structure provided by the tangent bundle. We do that by introducing a metric structure, where the horizontal part is isotropic and the generalized dynamics of the metric comes from the vertical part of the metric. For this, we consider a tangent bundle equipped with the metric tensor
\begin{equation}
\Gd = -\mathrm{d}t\otimes\mathrm{d}t  + a^2(t)\delta_{\kappa  \lambda}\,\mathrm{d}x^{\kappa}\otimes\mathrm{d}x^{\lambda} + h_{ab}(x,y)\,\delta y^a \otimes \delta y^b \label{frw on tm}
\end{equation}
The h-metric in rel.\eqref{frw on tm},
\begin{equation}
g_{ij} = diag\left(-1,a^2(t),a^2(t),a^2(t)\right)
\end{equation}
is the spatially flat FRW metric that depends only on the position on the base manifold. We study the case of an holonomic basis, i.e. the curvature coefficients of the nonlinear connection defined in rel.\eqref{Omega} are set to zero. Connection structure on this space is defined by coefficients given in rel.(\ref{metric d-connection 1}-\ref{metric d-connection 4}).

\subsection{Field equations}
Connection coefficients from rel.\eqref{metric d-connection 1} for the metric given in rel.\eqref{frw on tm} are reduced to
\begin{equation}
L^i_{jk} = \frac{1}{2}g^{ih}\left(\partial_k g_{hj} + \partial_j g_{hk} - \partial_h g_{jk} \right)
\end{equation}
These are just the Christoffel symbols for the h-metric $g_{ij}$. The Ricci curvature tensor components from rel.\eqref{d-ricci 1} read:
\begin{equation}
R_{ij} = \partial_kL^k_{ij} - \partial_jL^k_{ik} + L^m_{ij}L^k_{mk} - L^m_{ik}L^k_{mj}
\end{equation}
Due to the fact that $L^i_{jk}$ is the Christoffel connection, those are identified as the components of the Ricci tensor used in GR (General theory of Relativity), with well-known components for our choice of metric.

A simple cosmological model occurs by considering the energy and momentum described by an ideal fluid
\begin{equation}
\overset{1}T_{ij} = \left(\rho+P\right)\,u_i \, u_j + Pg_{ij}
\end{equation}
where $\rho (x)$ is the spacetime-dependent (isotropic) energy density, $P(x)$ is the spacetime-dependent (isotropic) pressure and $u$ is the four-velocity field of the fluid. Field equations from rel.\eqref{g field equation} are written as
\begin{equation}
R_{ij} - \frac{1}{2}\,g_{ij}\,R - \frac{1}{2}\,g_{ij}\,S = \kappa \overset{1}T_{ij} \label{ideal fluid}
\end{equation}
which reduce to ordinary differential equations for the scale factor $a(t)$ in the usual manner
\begin{gather}
\left( \frac{\dot a}{a} \right)^2 = \frac{\kappa}{3}\rho - \frac{1}{6}S \label{generalized friedmann equation 1} \lin
\frac{\ddot a}{a} = -\frac{\kappa}{6} \left(\rho+3P\right) - \frac{1}{6}S \label{generalized friedmann equation 2}
\end{gather}
where a dot denotes differentiation with respect to coordinate time. Relations \eqref{generalized friedmann equation 1}, \eqref{generalized friedmann equation 2} are the generalized Friedmann equations for this model. By considering these equations it follows that there is an equivalence with the equations of classical FRW model in their form (for $S=-2\Lambda$) but their dynamical evolution is different. In classical FRW model, homogeneous and isotropic constant $\Lambda$ is added ad hoc, while in our case, $S$-curvature is a dynamical anisotropic cosmological parameter which emerges from the additional degrees of freedom of the geometrical structure and plays the role of an effective cosmological constant.

At the present stage of cosmological evolution, $S$-curvature's value should be very small, so it could be described by a weak field framework. From rel.\eqref{ricci scalar s first order} we see that the dynamics of $S$ is connected with the weak field $\tilde h^{ab}(x,y)$ which is determined by the field equation \eqref{field equation 2 weak}. The matter source $\overset{2}{T}_{ab}$ must be chosen such that $S$-curvature's present value agrees with current cosmological observations.

It is known that in general relativity, the gravitational field is described by the metric tensor $g_{ij}(x)$. In our model, the gravitational field is described by two metric tensors $g_{ij}(x)$ and $h_{ab}(x,y)$ of which the dynamical evolution is connected. Consequently, the v-metric gives more degrees of freedom to the model and generalizes anisotropically the cosmological evolution (rel.\eqref{generalized friedmann equation 1} and \eqref{generalized friedmann equation 2}).

\subsection{Raychaudhuri equation of the model}\label{sec: raychaudhuri}

Given a four velocity field
\begin{equation}\label{4-velocity}
u^i(x) = \frac{dx^i}{d\tau}
\end{equation}
and setting $y^a(x) = \delta^a_i u^i(x)$, we get from rel.\eqref{bundle geodesic} the geodesics equation:
\begin{equation}
\frac{du^i}{d\tau} + 2G^i(x,u) = 0 \label{geo u}
\end{equation}
with $\tau$ being the proper time parameter defined in rel.\eqref{proper time}. The tangent vector to geodesics equation, rel.\eqref{geo u}, is given by the semispray field \cite{vacaru-stavrinos-gaburov-gonta 2006,bucataru-miron 2007}:
\begin{equation}
Y(x,u) \equiv u^i\partial_i - 2G^a(x,u)\pdot{a} \label{semispray}
\end{equation}
By using rel.\eqref{delta x}, equivalently on the adapted basis we get:
\begin{equation}
Y(x,u) = u^i\delta_i + \big( u^l N^a_l(x,u) - 2G^a(x,u)\big)\pdot{a}
\end{equation}

In the framework of Finslerian extensions, Raychaudhuri equations and energy conditions have been studied in \cite{stavrinos 2012,minguzzi 2015,stavrinos-alexiou 2016,singh-chaubey-singh 2015}. Raychaudhuri equation was developed on the tangent bundle of a spacetime and has been derived for a timelike congruence in \cite{stavrinos 2012}. Adapting a tangent field $Y$ on a timelike geodesic congruence, the h-space Raychaudhuri equation gives:
\begin{equation}
\dot \theta = -\frac{1}{3}\theta^2 - \sigma^2 + \omega^2 - R_{ij}u^iu^j + u^i \mathcal T^l_{ik} \overset{(h)}{\D}_l Y^k + u^i\Omega^a_{ik}\overset{(v)}{\D}_a Y^k \label{raychaudhuri bundle}
\end{equation}
where $\theta \equiv \overset{(h)}{\D}_k Y^k$ is the expansion, $\sigma_{ij} \equiv \overset{(h)}{\D}_{(i}Y_{j)} - \frac{1}{3}\theta P_{ij}$ is the shear and $\omega_{ij} \equiv \overset{(h)}{\D}_{[i} Y_{j]}$ is the vorticity of the congruence. We use the definitions $\sigma^2 \equiv \sigma_{ij}\sigma^{ij}$, $\omega^2 \equiv \omega_{ij}\omega^{ij}$ and the projection tensor is $P_{ij} \equiv g_{ij} + u_iu_j$. A dot denotes differentiation with respect to $\tau$.

For the case where $ \mathcal T^i_{jk} = 0 $ and because $ Y^k$ is a function of position $x$ only, equation \eqref{raychaudhuri bundle} gives
\begin{equation}
\dot \theta = -\frac{1}{3}\theta^2 - \sigma^2 - R_{ij}u^iu^j \label{raychaudhuri bundle 2}
\end{equation}
Field equations given in rel.\eqref{h field equation} can be manipulated to the form:
\begin{equation}
R_{ij} = \kappa \left( T_{ij} - \frac{1}{2}g_{ij}T\right) - \frac{1}{2}g_{ij}S \label{h field alternative}
\end{equation}
On the other hand, strong energy conditions for matter are defined as:
\begin{equation}
T_{ij}u^iu^j \geq \frac{1}{2}g_{ij}u^iu^jT
\end{equation}
which applied to rel.\eqref{h field alternative} give
\begin{equation}
R_{ij}u^iu^j \geq \frac{1}{2}S 
\end{equation}
where we used the normalization condition $u_iu^i = -1$. From the last relation and because of $\sigma^2 \geq 0$, rel.\eqref{raychaudhuri bundle 2} gives:
\begin{equation}
\dot \theta \leq -\frac{1}{3}\theta^2 -\frac{1}{2}S \label{raychaudhuri inequality}
\end{equation}
From relations (\ref{generalized friedmann equation 2}) and \eqref{raychaudhuri inequality} we notice that v-space scalar curvature $S$ can allow an increasing expansion $\theta$. The geometrical interpretation of $\theta = \overset{(h)}{\D}_k Y^k$ describes the deviation of neighbouring geodesics in a congruence along the direction of $ Y $. The proper time derivative of $\theta$ gives a measure of the relative acceleration of nearby test particles freely falling along the geodesics congruence. The inequality above sets an upper limit for the proper time derivative of $\theta$, which would necessarily be non-positive in the case of $S=0$ (as is the case of Riemannian Geometry) so nearby test particles would not accelerate relative to each other following just the geometry of spacetime. In our case, the upper limit also depends on $S$, so an acceleration of nearby test particles is possible.

\subsection{Energy conditions and cosmological bounce}
Ordinary matter fields, i.e. cold dark matter and radiation, obey certain energy conditions. In standard FRW cosmology, those matter fields are described as ideal fluids and are characterized by spatially homogeneous energy density $\rho$ and spatially homogeneous and isotropic pressure $P$ \cite{carroll 2004}. In this case, the weak, null and strong energy conditions, hereafter WEC, NEC and SEC respectively, are given as:
\begin{itemize}
\item WEC: $\rho\geq 0, \; \rho + P \geq 0 $
\item NEC: $\rho + P \geq 0 $
\item SEC: $\rho + P \geq 0, \; \rho + 3P \geq 0 $
\end{itemize}
from which, for the field equations given in relations \eqref{generalized friedmann equation 1} and \eqref{generalized friedmann equation 2}, we get
\begin{itemize}
\item WEC for generalized FRW: $\left( \dfrac{\dot a}{a}\right)^2 \geq -\dfrac{1}{6}S$ and $\left( \dfrac{\dot a}{a}\right)^2 \geq \left( \dfrac{\ddot a}{a}\right)$
\item NEC for generalized FRW: $\left( \dfrac{\dot a}{a}\right)^2 \geq \left( \dfrac{\ddot a}{a}\right)$
\item SEC for generalized FRW: $\left( \dfrac{\dot a}{a}\right)^2 \geq \left( \dfrac{\ddot a}{a}\right)$ and $ \left( \dfrac{\ddot a}{a}\right) \leq -\dfrac{1}{6}S $
\end{itemize}

A cosmological bounce on an FRW universe occurs when the conditions $\dot a(t_0) = 0$, $\ddot a(t_0) > 0$ are met for a coordinate time $t_0$. Studies of cosmological bounce in modified gravity have been made e.g. in \cite{singh-chaubey-singh 2015}. Subtracting rel.\eqref{generalized friedmann equation 1} from rel.\eqref{generalized friedmann equation 2} and applying the bounce conditions gives
\begin{equation}
\rho + P < 0 \label{energy condition violation}
\end{equation}
provided that $a>0$ and $\kappa > 0$. It is apparent that a cosmological bounce for this model requires the violation of all the aforementioned conditions.

\subsection{Scalar field}
Various scalar field models are used in cosmology in order to describe the accelerating expansion during the inflationary period of the universe. There have been studies of scalar field models in the framework of generalized geometry e.g. in \cite{stavrinos-ikeda 1999,stavrinos-ikeda 2000}.

In this section, we consider an anisotropic scalar field $\phi(x,y)$ on the tangent Lorentz bundle with a Lagrangian density
\begin{equation}
\mathcal L_\phi = -\frac{1}{2} \, \mathcal D_A \phi \, \mathcal D^A \phi - V(\phi) \label{scalar field lagrangian density}
\end{equation}
where $ \mathcal D_A \phi \, \mathcal D^A \phi =  \overset{(h)}{\mathcal D_i}\phi \, \overset{(h)}{\mathcal D^i}\phi + \overset{(v)}{\mathcal D_a}\phi\, \overset{(v)}{\mathcal D^a}\phi$. This is a direct generalization of the definition of a scalar field in regular $4-$dimensional spacetime. For a $d-$metric rel.(\ref{bundle metric}) this density is written as:
\begin{equation}
\mathcal L_\phi = -\frac{1}{2} \, g^{ij}\, \delta_i\phi \, \delta_j\phi - \frac{1}{2}h^{ab}\, \pdot{a}\phi \, \pdot{b}\phi - V(\phi)
\end{equation}
Definitions in relations \eqref{h-EM tensor} and \eqref{v-EM tensor} give:
\begin{align}
\overset{1}T_{ij}^{(\phi)} & \, = \delta_i\phi \, \delta_j\phi - g_{ij}\left( \frac{1}{2}\, g^{kl}\, \delta_k\phi \, \delta_l\phi + \frac{1}{2}\, h^{ab} \, \pdot{a}\phi\, \pdot{b}\phi + V(\phi)\right) \label{scalar energy-momentum 1} \lin
\overset{2}T_{ab}^{(\phi)} & \, = \pdot{a}\phi \, \pdot{b}\phi - h_{ab}\left( \frac{1}{2}\, g^{ij}\, \delta_i\phi \, \delta_j\phi + \frac{1}{2}\, h^{cd} \, \pdot{c}\phi\, \pdot{d}\phi + V(\phi)\right) \label{scalar energy-momentum 2}
\end{align}
Equations of motion for $\phi$ on the tangent bundle are given by extremalizing the action $S = \int \sqrt{\det [\mathcal G]}\,\mathcal L_\phi \, d^8\mathcal X$ for variations of $\phi$ or equivalently by the generalized Euler-Lagrange equations:
\begin{equation}
\pder{\mathcal L_\phi}{\phi} - \mathcal D_A\pder{\mathcal L_\phi}{\left(\mathcal D_A\phi\right)} = 0
\end{equation}
These give:
\begin{equation}
\overset{(h)}{\square}\phi + \overset{(v)}{\square}\phi - V' = 0 \label{generalized klein-gordon}
\end{equation}
where we denoted
\begin{equation}\label{d'alembertian}
\overset{(h)}{\square} \equiv g^{ij}\,\overset{(h)}{\mathcal D_i}\, \overset{(h)}{\mathcal D_j} \qquad \overset{(v)}{\square} \equiv h^{ab}\,\overset{(v)}{\mathcal D_a}\, \overset{(v)}{\mathcal D_b}
\end{equation}
the prime standing for differentiation with respect to $\phi$. Equation \eqref{generalized klein-gordon} is another generalization of the Klein-Gordon equation of the standard model of particle physics (provided that $V(\phi) = m^2 \phi^2/2\hbar^2$), along with rel.\eqref{klein gordon generalized}.

If we consider the specific case where the metric takes the form \eqref{frw on tm}, the connection coefficients \eqref{metric d-connection 1} reduce to the Christoffel coefficients for FRW metric. In this case, we take all the quantities on the bundle to be spatially homogeneous functions, so the functions $\phi(t,y)$, $N^a_i(t,y)$ depend on $t$ and $y$. For simplicity, we only take the $N^a_0$ coefficients of the nonlinear connection to be nonzero. From \eqref{d'alembertian} we get:
\begin{equation}
\overset{(h)}{\square}\phi = -\ddot \phi - 3H \dot \phi - \left( N^b_0\pdot{b}N^a_0 - \dot N^a_0\right) \pdot{a}\phi + 2N^a_0\pdot{a}\dot \phi - N^a_0N^b_0\pdot{a}\pdot{b}\phi + 3HN^a_0\pdot{a}\phi
\end{equation}
Equation \eqref{generalized klein-gordon} then gives:
\begin{align}
\ddot \phi + 3H \dot \phi + V' = & \, \overset{(v)}{\square}\phi - \left( N^b_0\pdot{b}N^a_0 - \dot N^a_0\right) \pdot{a}\phi + 2N^a_0\pdot{a}\dot \phi \nonumber \lin
& \, - N^a_0N^b_0\pdot{a}\pdot{b}\phi + 3HN^a_0\pdot{a}\phi \label{k-g frw}
\end{align}
where $H=\dot a/a$ is the Hubble parameter. For a v-metric given in rel.\eqref{v metric} and a Cartan-type connection, nonlinear connection components are given by rel.\eqref{N weak}. Then rel.\eqref{k-g frw} to first order with respect to $\tilde h_{ab}$ becomes:
\begin{align}
\ddot \phi + 3H \dot \phi + V' = & \, \eta^{ab}\pdot{a}\pdot{b}\phi - \eta^{ab}\eta^{cd}\left( \pdot{a} \tilde h_{bd} + \pdot{b} \tilde h_{ad} - \pdot{d} \tilde h_{ab}\right) \nonumber \lin
& \, + \frac{1}{2}y_b\, \ddot{\tilde{h}}^{ab} \pdot{a}\phi + y_b\, \dot{\tilde{h}}^{ab} \pdot{a}\dot \phi + \frac{3}{2}Hy_b\, \dot{\tilde{h}}^{ab} \pdot{a}\phi
\end{align}
We can model the scalar field as an ideal f\mbox{}luid by comparing relations \eqref{scalar energy-momentum 1} and \eqref{ideal fluid}. We get
\begin{equation}
\delta_i \phi = \sqrt{\rho_\phi + P_\phi} \, U_i \label{del f}
\end{equation}
and
\begin{equation}\label{fluid scalar 2}
P_\phi = -\frac{1}{2}\, g^{kl}\, \delta_k\phi \, \delta_l\phi - \frac{1}{2} \,h^{ab} \, \pdot{a}\phi\, \pdot{b}\phi - V(\phi)
\end{equation}
By combining the above two equations and solving for $\rho_\phi$ we find
\begin{equation}\label{fluid scalar 1}
\rho_\phi = -\frac{1}{2}\, g^{kl}\, \delta_k\phi \, \delta_l\phi + \frac{1}{2} \,h^{ab} \, \pdot{a}\phi\, \pdot{b}\phi + V(\phi)
\end{equation}
For the case of the generalized FRW metric rel.(\ref{frw on tm}) and for spatially homogeneous functions $\phi$ and $N^a_i$, energy density and pressure for the scalar f\mbox{}luid become:
\begin{equation}\label{frw scalar 1}
\rho_\phi = \frac{1}{2}\dot \phi^2 + \frac{1}{2}\, N^a_0N^b_0\pdot{a}\phi\, \pdot{b}\phi - \dot{\phi}N^a_0\pdot{a}\phi + \frac{1}{2}h^{ab}\pdot{a}\phi\, \pdot{b}\phi + V(\phi)
\end{equation}
and
\begin{equation}\label{frw scalar 2}
P_\phi = \frac{1}{2}\dot \phi^2 + \frac{1}{2}\, N^a_0N^b_0\pdot{a}\phi\, \pdot{b}\phi - \dot{\phi}N^a_0\pdot{a}\phi - \frac{1}{2}h^{ab}\pdot{a}\phi\, \pdot{b}\phi - V(\phi)
\end{equation}
For the case of the weak-field v-metric given in rel.\eqref{v metric}, energy density and pressure in relations \eqref{frw scalar 1} and \eqref{frw scalar 2} are to first order with respect to $\tilde h_{ab}$ calculated as:
\begin{align}
\rho_\phi = \frac{1}{2}\dot \phi^2 - \frac{1}{2}y_b\, \dot{\tilde{h}}^{ab} \dot\phi \, \pdot{a}\phi + \frac{1}{2}h^{ab}\pdot{a}\phi\, \pdot{b}\phi + V(\phi) \label{frw weak scalar 1} \lin
P_\phi = \frac{1}{2}\dot \phi^2 - \frac{1}{2}y_b\, \dot{\tilde{h}}^{ab} \dot\phi \, \pdot{a}\phi - \frac{1}{2}h^{ab}\pdot{a}\phi\, \pdot{b}\phi - V(\phi) \label{frw weak scalar 2}
\end{align}
Scalar field provides a viable model for a cosmological bounce. Going back to the bounce condition in rel.\eqref{energy condition violation}, relations \eqref{frw scalar 1} and \eqref{frw scalar 2} give:
\begin{equation}\label{scalar bounce condition}
\rho_\phi + P_\phi < 0 \Leftrightarrow \dot \phi^2 + N^a_0N^b_0\pdot{a}\phi\, \pdot{b}\phi - 2\dot{\phi}N^a_0\pdot{a}\phi < 0
\end{equation}
For the weak-field case and given relations \eqref{frw weak scalar 1} and \eqref{frw weak scalar 2} we get the necessary condition for a bouncing universe:
\begin{equation}\label{scalar weak bounce condition}
\rho_\phi + P_\phi < 0 \Rightarrow y_b\, \dot{\tilde h}^{ab} \dot \phi \, \pdot{a}\phi > 0
\end{equation}
We remark that the scalar field would not be able to provide a viable model for cosmological bounce in ordinary GR. To satisfy the bounce conditions in the weak field it is necessary to have a nonzero $\dot{\tilde h}^{ab}$ and also a scalar field with nonzero directional derivative $\pdot{a}\phi$, as we can see from rel.\eqref{scalar weak bounce condition}.

\section{Electromagnetic field tensor}\label{sec: em field tensor}

Previous studies that incorporate the electromagnetic field tensor and the associated Maxwell equations in the framework of the metric tangent bundle have been made, for example in \cite{stavrinos-diakogiannis 2004,pfeifer-wohlfarth 2011,voicu 2011,ikeda 1990,kouretsis 2014}. In the present work we study the electromagnetic field in the framework of tangent bundle's geometry. In this approach we assume that the vector potential $A$ is horizontal and isotropic:
\begin{equation}
A=A_i(x) \,\mathrm d x^i
\end{equation}
Isotropy here means that $A$ depends only on position coordinates $x^i$. Taking into account \cite{stavrinos-diakogiannis 2004} page 277, we consider symmetric connection coefficients $ L^h_{ij} $ and an isotropic $A_i(x)$ and we get a generalized description of the electromagnetic field tensor $F_{ij}=\partial_j A_i - \partial_i A_j$ on the tangent bundle:
\begin{align}\label{F st diak}
\tilde F_{ij} = \overset{(h)}{\mathcal D}_j A_i(x) - \overset{(h)}{\mathcal D}_i A_j(x) & = \delta_j A_i(x) - \delta_i A_j(x) - L^h_{ij}A_h + L^h_{ji}A_h(x) \nonumber\lin
& = \left( \partial_j - N^a_j\pdot{a}\right) A_i(x) - \left( \partial_i - N^a_i\pdot{a}\right) A_j(x)
\end{align}
or
\begin{equation}\label{F tilde}
\tilde F_{ij} = \partial_j A_i(x) - \partial_i A_j(x) = F_{ij}
\end{equation}
since
\begin{equation}\label{pdota=0}
\pdot{a} A_i(x) = 0
\end{equation}
From relations \eqref{F tilde} and \eqref{pdota=0} we get
\begin{align}
\frac{1}{2} \tilde F_{ij}(x)\, \mathrm{d}x^i \wedge \mathrm{d}x^j & = \partial_i A_j(x)\,\mathrm{d}x^i \wedge \mathrm{d}x^j \nonumber\lin
& = \delta_i A_j(x)\,\mathrm{d}x^i \wedge \mathrm{d}x^j
\end{align}
This is equivalent to the relation
\begin{equation}\label{FdA}
\tilde F = \mathrm d A
\end{equation}
where
\begin{equation}
\tilde F = \frac{1}{2}\tilde F_{ij}(x)\, \mathrm{d}x^i \wedge \mathrm{d} x^j
\end{equation}
From rel.\eqref{FdA} we get
\begin{equation}
\mathrm{d} \tilde F = 0 \label{1st maxwell equation}
\end{equation}

On the adapted basis, equation \eqref{1st maxwell equation} gives
\begin{equation}
\frac{1}{2}\delta_k \tilde F_{ij}\, \mathrm{d}x^k \wedge \mathrm{d}x^i \wedge \mathrm{d} x^j
\end{equation}
or
\begin{equation}
\partial_{[k}\tilde F_{ij]} = 0
\end{equation}
where the fact that the field tensor depends only on $x$ was used.

For the second part of Maxwell's equations, we consider the straightforward generalization:
\begin{equation}\label{maxwell equations covariant derivative}
\overset{(h)}{\mathcal D}_j \tilde F^{ij} = J^i
\end{equation}
where $ J = J^i(x,y) \delta_i$ is the electromagnetic $4-$current density. For a canonical $d-$connection we get:
\begin{align}\label{maxwell 2}
\overset{(h)}{\mathcal D}_j \tilde F^{ij} & = \partial_j \tilde F^{ij} + L^i_{kj} \tilde F^{kj} + L^j_{kj} \tilde F^{ik} \nonumber\lin
& = \partial_j \tilde F^{ij}(x) + \frac{1}{2} g^{jl}(x,y)\left( \delta_k g_{jl}(x,y) + \delta_j g_{kl}(x,y) - \delta_l g_{kj}(x,y)\right) \tilde F^{ik}(x) \nonumber\lin
& = \partial_j \tilde F^{ij}(x) + \frac{1}{2}g^{kl}(x,y)\delta_j g_{kl}(x,y) \tilde F^{ij}(x) = J^i(x,y)
\end{align}
This is equivalent to the relation
\begin{equation}
\mathrm{d} \star \tilde F = \star J \label{2nd maxwell equation}
\end{equation}
where $\star$ is the Hodge duality operator of the horizontal subspace.
To prove this, we first need the derivative of h-metric's $g_{ij}$ determinant with respect to $x^j$. For that we use the identity for a square matrix $M$:
\begin{equation*}
\ln(\det M) = tr(\ln M)
\end{equation*}
and find
\begin{equation}\label{metric determinant x derivative}
\delta_j \det[g] = \det[g]\, g^{kl}\delta_j g_{kl}
\end{equation}

Levi-Civita tensor for the h-space is
\begin{equation}\label{levi civita}
\epsilon_{ijkl} = \sqrt{-\det[g]}\, \overline{\epsilon}_{ijkl}
\end{equation}
where $\overline{\epsilon}_{ijkl}$ is the Levi-Civita symbol of the h-space, the convention $\overline{\epsilon}_{0123} = 1$ is followed.

Relation \eqref{2nd maxwell equation} in the adapted basis is written
\begin{equation}
\frac{1}{4}\delta_p \left( \epsilon_{ijkl} \tilde F^{ij}\right)\mathrm{d}x^p \wedge \mathrm{d}x^k \wedge \mathrm{d}x^l = \frac{1}{3!}J_i\, \epsilon^i_{jkl}\,\mathrm{d}x^j \wedge \mathrm{d}x^k \wedge \mathrm{d}x^l 
\end{equation}
Using relation \eqref{levi civita} we get:
\begin{equation}
\frac{1}{4}\,\overline{\epsilon}_{ijkl}\, \delta_p \left(\sqrt{-\det[g]}\,\tilde F^{ij}\right)\mathrm{d}x^p \wedge \mathrm{d}x^k \wedge \mathrm{d}x^l = \frac{1}{3!}\sqrt{-\det[g]}\, J_i\, \overline{\epsilon}^i_{jkl}\,\mathrm{d}x^j \wedge \mathrm{d}x^k \wedge \mathrm{d}x^l 
\end{equation}
After straightforward calculations this gives
\begin{equation}
\delta_j \left(\sqrt{-\det[g]}\,\tilde F^{ij}\right) = \sqrt{-\det[g]}\, J^i
\end{equation}
Finally, using rel.\eqref{metric determinant x derivative} we get the equation:
\begin{equation}\label{maxwell equation 2 coefficients}
\partial_j \tilde F^{ij}(x) + \frac{1}{2}g^{kl}(x,y)\delta_jg_{kl}(x,y)\tilde F^{ij}(x) = J^i(x,y)
\end{equation}
This concludes the proof that relations \eqref{maxwell equations covariant derivative} and \eqref{2nd maxwell equation} are equivalent.

We observe that equations \eqref{1st maxwell equation} and \eqref{2nd maxwell equation} of electromagnetism in our space are equivalent in form  with the Riemannian ones which is a result of our initial assumptions. Of course, the dynamics of the fields is not equivalent in the two geometries. The extended geometrical structure of the tangent bundle can result to an extension of physical predictions of the theory.

In our approach we have considered the 4-current density $J^i(x,y)$ to depend on $y$ in order to be consistent with the anisotropic geometric structure which is obvious in rel. \eqref{maxwell 2} and \eqref{maxwell equation 2 coefficients}. This can remove possible inconsistencies that could be inherited from the assumption of an isotropic field on an anisotropic background.

Another point of view that will give us isotropic field equations for electromagnetism and resolve any possible inconsistencies can be considered by introducing the action
	\begin{equation}
	S_{EM} = \int_M d^4 x \left( \int_{T_xM} d^4 y \sqrt{\det [G]} \mathcal L_{EM} \right)
	\end{equation}
	and perform the integration over the fiber before extremalizing over a subset of the base manifold $M$.

\section{Concluding remarks}\label{sec: conclusion}

In this work, we studied a weak-field model on the tangent bundle, which provides an insight in local anisotropy of modified Einstein gravity. Field equations rel.(\ref{field equation 1 weak}) extend Einstein's equations of GR with extra terms and a generalized energy-momentum tensor. The extra terms are connected with the anisotropic part of the geometry and can be interpreted as a possible anisotropy of the universe.

We derived the mass shell equation for a locally flat background h-metric, which is a generalization of the known $p^ip_i = -m^2$, rel.\eqref{generalized mass shell},\eqref{generalized mass shell final}. An extension of the Klein-Gordon equation was given and a dispersion relation for the scalar field's modes was derived as well, rel.\eqref{dispersion relation}.

A profound result of modified gravity theory on $TM$ is given in rel.\eqref{generalized friedmann equation 1},\eqref{generalized friedmann equation 2}, where the simple case of an isotropic FRW h-metric is considered. Field equations resemble the Friedmann equations, with an extra term being the scalar curvature $S$ of v-space. This extra term can give rise to an accelerating expansion of space. From a cosmological point of view, this can give an insight on a possible physical interpretation of v-metric $h_{ab}$.

On the other hand, it turns out that this generalized FRW cosmology cannot describe a bouncing universe, at least not for an ideal fluid matter field obeying basic energy conditions. We introduced a scalar field model and we derived the condition which can describe a bouncing cosmology, violating several energy conditions in the process. As well, a generalized form of Raychaudhuri equation was given in sec. \ref{sec: raychaudhuri} for the present model. In this approach, extra anisotropic terms can determine the accelerating/decelerating expansion of the universe.

Finally, we incorporated the electromagnetic field equations in the tangent bundle framework, taking the $4-$potential to be isotropic, and we resulted in equations similar to those of regular Riemannian models of gravity. Relations \eqref{1st maxwell equation} and \eqref{2nd maxwell equation} remain invariant in form under the introduction of local anisotropy at the description of gravity.

A further investigation of the consideration presented in this work regarding the electromagnetic field is required. This can be studied in the near future.

In the present work, the weak field equations provide an infrastructure for the study of gravitational waves and cosmological perturbations on the tangent Lorentz bundle. This will be a task for a future project.

\acknowledgements{
The authors would like to express their thanks to the unknown referees for their valuable
comments on the text. We also thank Dr S Basilakos for discussions on the manuscript.
}

\appendix

\section{Field equations on the tangent bundle}\label{sec: field equations}

In the following we consider a tangent Lorentz bundle equipped with a nonlinear connection, a $d-$metric, rel.\eqref{bundle metric}, and a canonical and distinguished $d-$connection, rel.(\ref{metric d-connection 1}-\ref{metric d-connection 4}).

The action of the fields in an arbitary closed subset $A$ of $TM$ is
\begin{equation}
S_{TM} = \dfrac{1}{2\kappa}S_{G} + S_{M}\equiv \dfrac{1}{2\kappa}\int_A\sqrt{|\Gdet|}\,\mathcal{L}_{G} \,d^8\mathcal X + \int_A\sqrt{|\Gdet|}\,\mathcal{L}_{M}\,d^8 \mathcal X \label{generalaction}
\end{equation}
where
\begin{equation}
S_{G} = \int_A d^8\mathcal{X}\sqrt{|\Gdet|}\,\R =  \int_A d^8\mathcal{X}\sqrt{|\Gdet|}\,(R+S) \label{bundle geometric action}
\end{equation}
is the gravitational part of the action and
\begin{equation}
S_{M} = \int_A d^8\mathcal{X}\sqrt{|\Gdet|}\,\mathcal{L}_M
\end{equation}
is the action of the matter fields, while $ \Gdet $ is the $d-$metric's determinant. This is a direct generalization of the Einstein-Hilbert action. Constant $\kappa$ will be specified by the limit of this framework where General Relativity is obtained. The volume element on $TM$ is defined by
\begin{equation}\label{volume element}
d^8\mathcal{X} = \mathrm{d}x^0 \wedge \mathrm{d}x^1 \wedge \mathrm{d}x^2 \wedge \mathrm{d}x^3 \wedge \delta y^0 \wedge \delta y^1 \wedge \delta y^2 \wedge \delta y^3
\end{equation}
We observe that the terms in $\delta y^a$ involving $\de x^i$, rel.\eqref{delta y}, will drop out in the exterior product with $ \mathrm{d}x^0 \wedge \mathrm{d}x^1 \wedge \mathrm{d}x^2 \wedge \mathrm{d}x^3 $, so we can equivalently write \eqref{volume element} as:
\begin{equation}\label{volume element 2}
d^8\mathcal{X} = \mathrm{d}x^0 \wedge \mathrm{d}x^1 \wedge \mathrm{d}x^2 \wedge \mathrm{d}x^3 \wedge \de y^0 \wedge \de y^1 \wedge \de y^2 \wedge \de y^3
\end{equation}

The independent fields of the underlying geometry are $g_{ij}(x,y)$, $h_{ab}(x,y)$ and $N^a_i(x,y)$. We will derive equations relating these fields to the matter fields by extremalizing the action rel.(\ref{generalaction}) with respect to variations $\delta g_{ij}$, $\delta h_{ab}$ and $\delta N^a_i$ which vanish at the boundary $\partial A$. The variation of the curvature coefficients $\Omega^a_{ij}$, rel.\eqref{Omega}, is given by
\begin{equation}
\delta\Omega^a_{ij} = \delta_j\delta N^a_i - \left(\pdot{b}N^a_i\right) \delta N^b_j - \delta_i\delta N^a_j + \left(\pdot{b}N^a_j\right)\delta N^b_i \label{omega variation}
\end{equation}
The variations of connection coefficients $ L^i_{jk} $ and $ C^k_{ia} $ are
\begin{align}
\delta L^i_{jk} = & - \frac{1}{2}\left(g_{jm}\overset{(h)}{\mathcal D}_k\,\delta g^{mi} + g_{km}\overset{(h)}{\mathcal D}_j\,\delta g^{mi} - g_{jm}g_{kh}\overset{(h)}{\mathcal D}{}^i\delta g^{mn}\right) \nonumber\lin
& -\left(C^i_{ja}\delta N^a_k + C^i_{ka}\delta N^a_j - g^{ir}g_{hk}\delta N^a_r\right) \label{L variation} \lin
\delta C^k_{ia} = &  \frac{1}{2}C_{ija}\delta g^{jk} - \frac{1}{2}C^k_{ma}g_{in}\delta g^{mn} - \frac{1}{2}g_{im} \overset{(v)}{\D}_a\delta g^{mk} \label{C variation} 
\end{align}
The variation of h-Ricci tensor $R_{ij}$ is given by
\begin{align}
\delta R_{ij} = \, & \overset{(h)}{\mathcal D}_k\left(\delta L^k_{ij} + C^k_{ia}\delta N^a_j\right) - \overset{(h)}{\mathcal D}_j\left(\delta L^k_{ik} + C^k_{ia}\delta N^a_k\right) + \overset{(v)}{\mathcal D}_a\left(\frac{1}{2}\Omega^a_{kj}g_{im}\delta g^{mk}\right) \nonumber\lin
& + \left[\overset{(h)}{\mathcal D}_j\,C^k_{ia} + C^k_{ib}L^b_{ja} - C^k_{ib}\left(\pdot{a}N^b_j\right) - \left(\pdot{a}L^k_{ij}\right)\right]\delta N^a_k \nonumber\lin
& - \left[\overset{(h)}{\mathcal D}_k\,C^k_{ia} + C^k_{ib}L^b_{ka} - C^k_{ib}\left(\pdot{a}N^b_k\right) - \left(\pdot{a}L^k_{ik}\right)\right]\delta N^a_j \nonumber\lin
& - \frac{1}{2}\left[\Omega^a_{kj}C^m_{ha}g_{im} - \Omega^a_{mj}C^m_{ha}g_{ik} + \left(\overset{(v)}{\mathcal D}_a\,\Omega^a_{kj}\right)g_{ih}\right]\delta g^{hk}
\end{align}
and the variation of the h-Ricci scalar tensor $R$ is given by
\begin{equation}
\delta R = \left(g^{ik}\pdot{a}L^j_{ij} - g^{ij}\pdot{a}L^k_{ij}\right)\delta N^a_k + \left(\Omega^a_{ki}C^k_{ja} + R_{ij}\right)\delta g^{ij} + \D_B Z^B \label{h curvature variation}
\end{equation}
where $\D_B \equiv \delta^i_B\overset{(h)}{\D}_i + \delta^a_B\overset{(v)}{\D}_a $ and $Z^B = (Z^k,0)$, with
\begin{equation*}
Z^k \equiv g^{ij}\delta L^k_{ij} - g^{ik}\delta L^j_{ij}
\end{equation*}
The variation of v-Ricci tensor $S_{ab}$ and v-Ricci scalar tensor $S$ takes the form
\begin{equation}
\delta S_{ab} = \overset{(v)}{\mathcal D}_c\,\delta C^c_{ab} - \overset{(v)}{\mathcal D}_b\,\delta C^c_{ac} \nonumber
\end{equation}
\begin{equation}
\delta S = S_{ab}\delta h^{ab} + \D_B K^B \label{v curvature variation}
\end{equation}
where $K^B = (0,K^c)$, with
\begin{equation}
K^c \equiv h^{ab}\delta C^c_{ab} -h^{ac}\delta C^b_{ab}
\end{equation}
Metric tensor, rel.\eqref{bundle metric}, is represented in the adapted basis as a block diagonal matrix. From a well known theorem regarding such matrices we get
\begin{equation}\label{block diagonal det}
\Gdet  = \det[g]\det[h]
\end{equation}
where $\det[g]$ and $\det[h]$ are the determinants of the h-metric and v-metric respectively. As far as the variation of the metric determinant's square root is concerned, we find
\begin{align}
\delta\sqrt{\Gdet} & = \delta\left(\sqrt{-\det[g]}\sqrt{-\det[h]}\right) \lin
& = \frac{1}{2\sqrt{-\det[g]}\sqrt{-\det[h]}}\big(\det[g]\,\delta\!\det[h] + \det[h]\,\delta\!\det[g]\big) \lin
& = -\frac{1}{2} \sqrt{\Gdet}
\left(g_{ij}\delta g^{ij} + h_{ab}\delta h^{ab}\right) \label{d-metric determinant variation}
\end{align}
From relations \eqref{h curvature variation}, \eqref{v curvature variation} and \eqref{d-metric determinant variation} we get the variation of the geometrical part of the action:
\begin{align}
\delta S_{G} = \int_A d^8\mathcal{X} & \sqrt{\Gdet}\, \left[ -\frac{1}{2}\left(g_{ij}\delta g^{ij} + h_{ab}\delta h^{ab}\right)(R+S) \right. \nonumber\lin 
& + \left(g^{ik}\pdot{a}L^j_{ij} - g^{ij}\pdot{a}L^k_{ij}\right)\delta N^a_k  \nonumber\lin
& \left. + \left(\Omega^a_{ki}C^k_{ja} + R_{ij}\right)\delta g^{ij} + S_{ab}\delta h^{ab} + \D_B\left(Z^B + K^B\right)\right] \label{delta SG}
\end{align}
From Stokes theorem we get:
\begin{equation}
\int_A d^8\mathcal{X}\,\D_B\left(\sqrt{\Gdet}\,(Z+K)^B\right) = \oint_{\partial A} d^7\mathcal{X}\sqrt{|\det [\mathscr{G}]|}\, n_B\, (Z+K)^B = 0 \label{action boundary}
\end{equation}
where $\det [\mathscr{G}]$ is the determinant of the metric $ \mathscr{G} $ of the boundary space $\partial A$. We have assumed the vanishing of the boundary term.

Extremization of the action $S_{TM}$ with respect to $g_{ij}$, $h_{ab}$ and $N^a_i$ gives:
\begin{align}
&\int_A d^8\mathcal{X}\,\sqrt{\Gdet}\,\frac{1}{2\kappa}\left[R_{ij} - \frac{1}{2}(R+S)g_{ij} + \Omega^a_{ki}C^k_{ja} - \kappa \overset{1}T_{ij}\right]\delta g^{ij} \nonumber\lin
+ &\int_A d^8\mathcal{X}\,\sqrt{\Gdet}\,\frac{1}{2\kappa}\left[S_{ab} - \frac{1}{2}(R+S)h_{ab} - \kappa \overset{2}T_{ab}\right]\delta h^{ab} \nonumber\lin
+ &\int_A d^8\mathcal{X}\,\sqrt{\Gdet}\,\frac{1}{2\kappa}\left(g^{ik}\pdot{a}L^j_{ij} - g^{ij}\pdot{a}L^k_{ij} - \kappa\overset{3}T{\,}{_a^k}\right)\delta N^a_k  = 0 \label{extremization in TM}
\end{align}
where the energy momentum tensor coefficients are defined as
\begin{align}
\overset{1}T_{ij} \equiv -\frac{2}{\sqrt{\Gdet}}\frac{\delta\left(\sqrt{\Gdet}\,\mathcal{L}_{M}\right)}{\delta g^{ij}} \label{Tij} \lin
\overset{2}T_{ab} \equiv -\frac{2}{\sqrt{\Gdet}}\frac{\delta\left(\sqrt{\Gdet}\,\mathcal{L}_{M}\right)}{\delta h^{ab}} \label{Tab} \lin
\overset{3}T{\,}{_a^k} \equiv -\frac{2}{\sqrt{\Gdet}}\frac{\delta\left(\sqrt{\Gdet}\,\mathcal{L}_M\right)}{\delta N^a_k}
\end{align}
From rel.\eqref{extremization in TM} we get the field equations
\begin{equation}
R_{(ij)} - \frac{1}{2}(R+S)g_{ij} + \Omega^a_{k(i}C^k_{j)a} = \kappa \overset{1}T_{ij} \label{hfe}
\end{equation}
\begin{equation}
S_{ab} - \frac{1}{2}(R+S)h_{ab} = \kappa \overset{2}T_{ab} \label{vfe}
\end{equation}
\begin{equation}
g^{ik}\pdot{a}L^j_{ij} - g^{ij}\pdot{a}L^k_{ij} = \kappa \overset{3}T{\,}{_a^k} \label{N field equation}
\end{equation}
Here we have presented a more general approach than the one we follow at the other sections. Specifically, we have assumed $g_{ij}$, $h_{ab}$ and $N^a_i$ to be independent dynamic fields on the tangent bundle. However, at the rest of our work we consider the nonlinear connection $N^a_i$ as an a priori defined structure on the tangent bundle, so field equation \eqref{N field equation} cannot be considered valid.

\subsection*{Determination of constant $\kappa$}
We consider now the limit where $N^a_i$ goes to zero, $h_{ab}$ goes to $\eta_{ab} = diag(-1,1,1,1)$ and differentiation with respect to $y$ of any quantity defined on $TM$ goes to zero. The adapted basis $\{\delta_i,\pdot{_a}\}$ and its dual $\{\mathrm{d}x^i,\delta y^a\}$ defined in relations (\ref{delta x}-\ref{delta y}) reduce then to $\{\partial_i,\pdot{_a}\}$ and $\{\mathrm{d}x^i,\mathrm{d}y^a\}$ respectively. Curvature coefficients $\Omega^a_{ij}$,  rel.(\ref{Omega}), vanish in this limit. The metric on the tangent bundle, rel.(\ref{bundle metric}), becomes:
\begin{equation}\label{bundle metric gr}
\mathcal{G} = g_{ij}(x)\,\mathrm{d}x^i \otimes \mathrm{d}x^j + \eta_{ab}\,\mathrm{d} y^a \otimes \mathrm{d} y^b 
\end{equation}
The connection coefficients $L^i_{jk}$ given in rel.\eqref{metric d-connection 1} are then
\begin{equation}
L^i_{jk} = \frac{1}{2}g^{ih}\left(\partial_kg_{hj} + \partial_jg_{hk} - \partial_hg_{jk}\right)
\end{equation}
We observe that $L^i_{kj}=\gamma^i_{kj}$ in the GR limit. The rest of the connection coefficients rel.(\ref{metric d-connection 2}-\ref{metric d-connection 4}) vanish. Moreover, the quantities $R_{ij}$ and $R$ defined in relations \eqref{d-ricci 1},\eqref{hv ricci scalar} are given in this limit as:
\begin{equation}\label{r1}
R_{ij} = \partial_k\gamma^k_{ij} - \partial_j\gamma^k_{ik} + \gamma^m_{ij}\gamma^k_{mk} - \gamma^m_{ik}\gamma^k_{mj}
\end{equation}
\begin{equation}\label{r2}
R=g^{ij}R_{ij}
\end{equation}
These are identified as the Ricci tensor and Ricci scalar of Riemannian geometry for the metric $g_{ij}$ as can be seen by the respective definitions, while $S_{ab}$ and $S$ vanish. Using rel.\eqref{block diagonal det}, the determinant of the metric of rel.(\ref{bundle metric gr}) is given as:
\begin{equation}
\Gdet = \det[g]\det[\eta] = -\det[g]
\end{equation}
so rel.\eqref{Tij} becomes
\begin{equation}\label{gr em}
\overset{1}T_{ij} \equiv -\frac{2}{\sqrt{-\det[g]}}\frac{\delta\left(\sqrt{-\det[g]}\,\mathcal{L}_{M}\right)}{\delta g^{ij}} 
\end{equation}
This is the definition of the energy momentum tensor of GR. Field equations rel.(\ref{hfe}) reduce to
\begin{equation}\label{gr fe}
R_{ij} - \frac{1}{2}R g_{ij} = \kappa \overset{1}T_{ij}
\end{equation}
Given the relations \eqref{r1}, \eqref{r2} and \eqref{gr em}, we identify these as the Einstein field equations for the metric $g_{ij}(x)$.

Geodesics equation \eqref{finsler geodesics} reduces to
\begin{equation}\label{gr geodesics}
\der{y^a}{\tau} + \gamma^a_{ij}(x)y^i y^j = 0,\qquad y^i=\frac{dx^i}{d\tau} 
\end{equation}
This is the geodesics equation defined in GR for a metric tensor $g_{ij}(x)$.

From relations \eqref{gr em}, \eqref{gr fe} and \eqref{gr geodesics} we deduce that in this limit we get ordinary general relativity for a spacetime equipped with metric tensor $g_{ij}(x)$. Constant $\kappa$ is thus specified as
\begin{equation}
\kappa=8\pi G/c^4
\end{equation}
where $G$ is the gravitational constant.

Field equations \eqref{vfe} in this limit become
\begin{equation}
-\frac{1}{2}R\,\eta_{ab} =  \kappa \overset{2}T_{ab}
\end{equation}
These equations restrict the way matter fields can depend upon the v-metric. In this sense, they do not affect the dymanics of $g_{ij}(x)$.

\section{Horizontal space Ricci curvature coefficients and Ricci scalar}\label{first order Ricci appendix}
The components of the h-Ricci tensor are obtained from relations \eqref{d-ricci 1} and \eqref{d-1 first order} as:
\begin{align}
R_{ij} = & \,\bcg R_{ij} + \big( \bcg R^r_{ikj} + \bcg\,\gamma^m_{kj}\,\bcg\,\gamma^r_{im} - \bcg\,\gamma^m_{km}\,\bcg\,\gamma^r_{ij} - \bcg\,\gamma^r_{ij}\partial_k\big)\tilde g^k_r \nonumber\lin
&\, + \delta^p_{(i}\delta^q_{j)} \left\{ \Big( \partial_k\bcg g^{kr}\Big) \Big( \partial_p\tilde g_{qr} - \frac{1}{2}\partial_r\tilde g_{pq}\Big) - \Big( \partial_p\bcg g^{kr}\Big) \partial_q\tilde g_{kr} \right. \nonumber\lin
&\, + \left. \bcg g^{kr} \Big[ \partial_p\partial_k\tilde g_{qr} - \frac{1}{2}\partial_k\partial_r\tilde g_{pq} - \frac{1}{2}\partial_p\partial_q\tilde g_{kr} + \frac{1}{2}\,\bcg\,\gamma^m_{pq}\partial_m\tilde g_{kr} \right. \nonumber\lin
&\, - \left. \bcg\,\gamma^m_{pk}\big( \partial_q\tilde g_{mr} + \partial_m \tilde g_{qr} - \partial_r \tilde g_{mq}\big) + \bcg\,\gamma^m_{km}\big( \partial_p\tilde g_{qr} - \frac{1}{2}\partial_r\tilde g_{pq}\big) \Big] \right\} \label{ricci r first order}
\end{align}
Using this result, the h-Ricci scalar curvature from rel.\eqref{hv ricci scalar} to first order with respect to $\tilde g_{ij}$ and $\tilde h_{ab}$ equals:
\begin{align}
R = & \, \bcg R - \tilde g^{ij}\,\bcg R_{ij} + \bcg g^{ij}\big( \bcg R^r_{ikj} + \bcg\,\gamma^m_{kj}\,\bcg\,\gamma^r_{im} - \bcg\,\gamma^m_{km}\,\bcg\,\gamma^r_{ij} - \bcg\,\gamma^r_{ij}\partial_k\big)\tilde g^k_r \nonumber\lin
& \, + \bcg g^{ij} \left\{ \Big( \partial_k\bcg g^{kr}\Big) \Big( \partial_i\tilde g_{jr} - \frac{1}{2}\partial_r\tilde g_{ij}\Big) - \Big( \partial_i\bcg g^{kr}\Big) \partial_j\tilde g_{kr} \right. \nonumber\lin
& \, + \left. \bcg g^{kr} \Big[ \partial_i\partial_k\tilde g_{jr} - \frac{1}{2}\partial_k\partial_r\tilde g_{ij} - \frac{1}{2}\partial_i\partial_j\tilde g_{kr} + \frac{1}{2}\,\bcg\,\gamma^m_{ij}\partial_m\tilde g_{kr} \right. \nonumber\lin
& \, - \left. \bcg\,\gamma^m_{ik}\big( \partial_j\tilde g_{mr} + \partial_m \tilde g_{jr} - \partial_r \tilde g_{mj}\big) + \bcg\,\gamma^m_{km}\big( \partial_i\tilde g_{jr} - \frac{1}{2}\partial_r\tilde g_{ij}\big) \Big] \right\} \label{ricci scalar r first order}
\end{align}

\section{Coeficients $\mathcal{A}_{ij}$ and $\mathcal{B}_{ab}$}\label{first order coefficients appendix}

The quantities $ \mathcal{A}_{ij} $ and $ \mathcal{B}_{ab} $ which appear in relations \eqref{field equation 1 weak} and \eqref{field equation 2 weak} respectively, are the terms which are of first order in $\tilde g_{ij}$ and $\tilde h_{ab}$ and are defined as follows:

\begin{align}
\mathcal{A}_{ij} = & - \frac{1}{2}\tilde g_{ij} \bcg R + \frac{1}{2}\,\bcg g_{ij}\tilde g^{pq} \,\bcg R_{pq} - \big(  \bcg\,\gamma^m_{km}\,\bcg\,\gamma^r_{ij} + \bcg\,\gamma^r_{ij}\partial_k \big)\tilde g^k_r \nonumber\lin
&+ \delta^p_{(i}\delta^q_{j)} \left\{ \left(\bcg R^r_{pkq} + \bcg\,\gamma^m_{kq}\,\bcg\,\gamma^r_{pm}\right)\tilde g_r^k + \Big( \partial_k\bcg g^{kr}\Big) \Big( \partial_p\tilde g_{qr} - \frac{1}{2}\partial_r\tilde g_{pq}\Big) - \Big( \partial_p\bcg g^{kr}\Big) \partial_q\tilde g_{kr} \right. \nonumber\lin
&+ \left. \bcg g^{kr} \Big[ \partial_p\partial_k\tilde g_{qr} - \frac{1}{2}\partial_k\partial_r\tilde g_{pq} - \frac{1}{2}\partial_p\partial_q\tilde g_{kr} + \frac{1}{2}\,\bcg\,\gamma^m_{pq}\partial_m\tilde g_{kr} \right. \nonumber\lin
&- \left. \bcg\,\gamma^m_{pk}\big( \partial_q\tilde g_{mr} + \partial_m \tilde g_{qr} - \partial_r \tilde g_{mq}\big) + \bcg\,\gamma^m_{km}\big( \partial_p\tilde g_{qr} - \frac{1}{2}\partial_r\tilde g_{pq}\big) \Big] \right\} \nonumber\lin
& - \frac{1}{2}\bcg g_{ij} \bcg g^{pq} \left\{ \Big( \partial_k\bcg g^{kr}\Big) \Big( \partial_p\tilde g_{qr} - \frac{1}{2}\partial_r\tilde g_{pq}\Big) - \Big( \partial_p\bcg g^{kr}\Big) \partial_q\tilde g_{kr} \right. \nonumber\lin
& + \left. \bcg g^{kr} \Big[ \partial_p\partial_k\tilde g_{qr} - \frac{1}{2}\partial_k\partial_r\tilde g_{pq} - \frac{1}{2}\partial_p\partial_q\tilde g_{kr} + \frac{1}{2}\,\bcg\,\gamma^m_{pq}\partial_m\tilde g_{kr} \right. \nonumber\lin
& - \left. \bcg\,\gamma^m_{pk}\big( \partial_q\tilde g_{mr} + \partial_m \tilde g_{qr} - \partial_r \tilde g_{mq}\big) + \bcg\,\gamma^m_{km}\big( \partial_p\tilde g_{qr} - \frac{1}{2}\partial_r\tilde g_{pq}\big) \Big] \right\} \nonumber\lin
& -\frac{1}{2} \bcg g_{ij} \bcg g^{pq}\big( \bcg R^r_{pkq} + \bcg\,\gamma^m_{kq}\,\bcg\,\gamma^r_{pm} - \bcg\,\gamma^m_{km}\,\bcg\,\gamma^r_{pq} - \bcg\,\gamma^r_{pq}\partial_k\big)\tilde g^k_r \nonumber\lin
& - \frac{1}{2} \bcg g_{ij}\left( \pdot{a}\pdot{b} \tilde h^{ab} - \overset{(v)}{\square} \tilde h\right) 
\end{align}

\begin{align}
\mathcal{B}_{ab} = &  \frac{1}{2}\left( \dot{\partial}^c\pdot{a} \tilde h_{bc} + \dot{\partial}^c\pdot{b} \tilde h_{ac} - \overset{(v)}{\square} \tilde h_{ab} - \pdot{a}\pdot{b} \tilde h\right) -\frac{1}{2}\tilde h_{ab}\bcg R \nonumber\lin
&+ \frac{1}{2} \eta_{ab}\tilde g^{ij}\,\bcg R_{ij} - \frac{1}{2}\eta_{ab}\bcg g^{ij}\big( \bcg R^r_{ikj} + \bcg\,\gamma^m_{kj}\,\bcg\,\gamma^r_{im} - \bcg\,\gamma^m_{km}\,\bcg\,\gamma^r_{ij} - \bcg\,\gamma^r_{ij}\partial_k\big)\tilde g^k_r \nonumber\lin
&-\frac{1}{2}\eta_{ab}\bcg g^{ij} \left\{ \Big( \partial_k\bcg g^{kr}\Big) \Big( \partial_i\tilde g_{jr} - \frac{1}{2}\partial_r\tilde g_{ij}\Big) - \Big( \partial_i\bcg g^{kr}\Big) \partial_j\tilde g_{kr} \right. \nonumber\lin
&+ \left. \bcg g^{kr} \Big[ \partial_i\partial_k\tilde g_{jr} - \frac{1}{2}\partial_k\partial_r\tilde g_{ij} - \frac{1}{2}\partial_i\partial_j\tilde g_{kr} + \frac{1}{2}\,\bcg\,\gamma^m_{ij}\partial_m\tilde g_{kr} \right. \nonumber\lin
&- \left. \bcg\,\gamma^m_{ik}\big( \partial_j\tilde g_{mr} + \partial_m \tilde g_{jr} - \partial_r \tilde g_{mj}\big) + \bcg\,\gamma^m_{km}\big( \partial_i\tilde g_{jr} - \frac{1}{2}\partial_r\tilde g_{ij}\big) \Big] \right\} \nonumber\lin
&- \frac{1}{2}\eta_{ab}\left( \pdot{a}\pdot{b} \tilde h^{ab} - \overset{(v)}{\square} \tilde h\right)
\end{align}

\section{Canonical momentum for the massive particle in a weakly anisotropic metric space}\label{app: massive particle momentum}
We consider a tangent bundle equipped with the metric tensor defined in rel.(\ref{total metric}-\ref{v metric}). The trajectory in spacetime for the massive particle of section \ref{subsec: mass shell} is described by the Lagrangian rel.(\ref{pp lagrangian first order}):
\begin{equation*}
L = -m \big(-y^i y_i \big)^{1/2} = -m \left( L_R - \frac{1}{2}L_R^{-1}\tilde g_{ij}(x,y) \, y^i y^j\right)
\end{equation*}
where $y^i y_i = g_{ij}y^i y^j$ and in the second equality only terms up to first order with respect to $\tilde g_{ij}$ are kept. The Riemannian norm $L_R$ is defined in rel.\eqref{Riemannian norm} as:
\begin{equation*}
L_R = \left( - \bcg g_{ij}(x) \, y^i y^j\right)^{1/2}
\end{equation*}
The canonical momentum is found as
\begin{equation}\label{canonical momentum full}
p_i = \pdot{i}L = mL_R^{-1} \left[ y_i + \frac{1}{2} \left( \pdot{i} \tilde g_{kl} +  L_R^{-2}y_i\tilde g_{kl}\right) y^k y^l \right]
\end{equation}
We normalize the fiber coordinates as $y^i y_i = - 1$. We find to first order with respect to $\tilde g_{ij}$:
\begin{gather}
-y^i y_i = L_R - \frac{1}{2}L_R^{-1}\tilde g_{ij}y^i y^j = 1 \nonumber\lin
\Rightarrow L_R^2 - L_R - \frac{1}{2}\tilde g_{ij}y^i y^j = 0
\end{gather}
Solving the quadratic equation with respect to $L_R$ gives to first order with respect to the perturbation:
\begin{equation}\label{Riemannian norm normalized}
L_R = \frac{1}{2} \big( 1 \pm \sqrt{1 + 2 \tilde g_{ij}y^i y^j}\big) = 1 + \frac{1}{2} \tilde g_{ij}y^i y^j
\end{equation}
where the second solution, namely $L_R = -\frac{1}{2}\tilde g_{ij}y^i y^j$ is rejected as it gives a vanishing Riemannian norm in zero order.

The inverse $L_R^{-1}$ of rel.\eqref{Riemannian norm normalized}, as well as $L_R^{-2}$, are easily found to be
\begin{equation}\label{Riemannian norm normalized inverse}
L_R^{-1} =1 - \frac{1}{2}\tilde g_{ij}y^i y^j, \quad L_R^{-2} = 1 - \tilde g_{ij}y^i y^j
\end{equation}

Putting together relations \eqref{canonical momentum full} and \eqref{Riemannian norm normalized inverse} and keeping terms up to first order with respect to $\tilde g_{ij}$ we get the canonical momentum given in rel.\eqref{4 momentum first order}:
\begin{equation}
p_i = m \left[ y_i + \frac{1}{2} \left(\pdot{i} \tilde g_{kl}\right) y^k y^l \right]
\end{equation}
\raggedbottom
\pagebreak


\end{document}